\documentclass[11pt]{article}
\usepackage{url}
\usepackage{booktabs} % tables
\usepackage{graphicx} % figures
\usepackage{subcaption} % subtable
\usepackage[labelfont=bf]{caption}
\usepackage{mathtools}
\usepackage{amsmath}
 \usepackage{amsthm}
\usepackage{amsfonts}
\usepackage{amssymb}
\usepackage{enumitem}
\usepackage{algorithm}
\usepackage[noend]{algpseudocode}
\usepackage{multicol}
\usepackage{multirow}
\usepackage{xspace}
\usepackage{color}
\usepackage{hyphenat}
\usepackage{xfrac} % better fractions
\usepackage{siunitx} % num command
\usepackage[capitalise]{cleveref} % smart references
\usepackage[square,numbers]{natbib} % better references

\newtheorem{definition}{Definition}
\newtheorem{observation}{Observation}
\newtheorem{problem}{Problem}
\newtheorem{example}{Example}

\newlist{todolist}{itemize}{2}
\setlist[todolist]{label=$\square$}

\newcommand{\spara}[1]{\smallskip\noindent{\bf #1}}
\newcommand{\mpara}[1]{\medskip\noindent{\bf #1}}

\newcommand{\vdist}[1]{\ensuremath{\delta_{#1}}\xspace}
\newcommand{\hdist}[1]{\ensuremath{\Delta_{#1}}\xspace}
\newcommand{\vhdist}[1]{\ensuremath{\mathsf{d}_{#1}}\xspace}
\newcommand{\vest}[1]{\ensuremath{\tilde{\delta_{#1}}}\xspace}
\newcommand{\hest}[1]{\ensuremath{\tilde{\Delta_{#1}}}\xspace}
\newcommand{\vhest}[1]{\ensuremath{\tilde{\mathsf{d}_{#1}}}\xspace}
\newcommand{\smax}{\ensuremath{s_{\textsf{max}}}\xspace}
\newcommand{\dmin}{\ensuremath{d_{\textsf{min}}}\xspace}
\newcommand{\algo}{\textsc{HypED}\xspace}
\newcommand{\base}{\textsc{BaseLine}\xspace}
\newcommand{\ccswise}{\textsc{CCS-SW}\xspace}
\newcommand{\ccsbase}{\textsc{CCS-IS}\xspace}
\newcommand{\lineg}{\textsc{LG}\xspace}
\newcommand{\ctl}{\textsc{CTL}\xspace}
\newcommand{\hl}{\textsc{HL}\xspace}
\newcommand{\samp}{\textsf{Sampling}\xspace}
\newcommand{\rank}{\textsf{RankAgg}\xspace}

\newcommand{\rev}[1]{\textcolor{black}{#1}}
\newcommand{\revv}[1]{\textcolor{black}{#1}}

\begin{document}
%\title{Approximating Hyper-distance Queries in Hypergraphs}
\title{Hyper-distance Oracles in Hypergraphs}

% for each affiliation group.
\author{Giulia~Preti\\
        Gianmarco~De~Francisci~Morales\\
        Francesco~Bonchi\\
        {\small CENTAI, Italy}
%
%\institute{
%    G. Preti, G. De Francisci Morales, and F. Bonchi \at
%
}

\date{}

\maketitle \sloppy

\begin{abstract}
We study point-to-point distance estimation in hypergraphs, where the query is parameterized by a positive integer $s$, which defines the required level of overlap for two hyperedges to be considered adjacent.
To answer $s$-distance queries, we first explore an oracle based on the line graph of the given hypergraph and discuss its limitations: the main one is that the line graph is typically orders of magnitude larger than the original hypergraph.

We then introduce \algo, a landmark-based oracle with a predefined size, built directly on the hypergraph, thus avoiding constructing the line graph.
Our framework allows to approximately answer vertex-to-vertex, vertex-to-hyperedge, and hyperedge-to-hyperedge $s$-distance queries for any value of $s$.
A key observation at the basis of our framework is that, as $s$ increases, the hypergraph becomes more fragmented.
We show how this can be exploited to improve the placement of landmarks, by identifying the $s$-connected components of the hypergraph.
For this task, we devise an efficient algorithm based on the union-find technique and a dynamic inverted index.
%Finally, we compare different techniques to determine the number of landmarks to assign to each $s$-connected component, and to select which hyperedges will act as landmarks for each component.

We experimentally evaluate \algo on several real-world hypergraphs and prove its versatility in answering $s$-distance queries for different values of $s$.
Our framework allows answering such queries in fractions of a millisecond, while allowing fine-grained control of the trade-off between index size and approximation error at creation time.
Finally, we prove the usefulness of the $s$-distance oracle in two applications, namely, hypergraph-based recommendation and the approximation of the $s$-closeness centrality of vertices and hyperedges in the context of protein-to-protein interactions.
\end{abstract}

% !TEX root = ../main.tex
\section{Introduction}\label{sec:intro}
Computing point-to-point shortest-path distances is a key primitive in network-structured data~\cite{goldberg2007point},
\rev{with widespread applications across various domains, including route planning~\cite{goldberg2006reach}, proximity search in databases~\cite{goldman1998proximity}, metabolic pathway analysis~\cite{rahman2005metabolic}, web search ranking~\cite{vieira2007efficient}, and path finding in social networks~\cite{kleinberg2000navigation}.}
\rev{The conventional approach to answering shortest-path queries involves using a single-source shortest-path algorithm for each query. However, this method leads to query times that grow linearly with the size of the network. An alternative approach involves an exhaustive computation of all possible queries through an all-pairs shortest-path algorithm, and storing these results in a lookup table. Yet, this method consumes a prohibitive amount of memory.}
In the context of real-world massive networks and with online applications requiring near-instantaneous point-to-point distances, these conventional approaches become impractical due to their prohibitively large query times or space requirements.
For this reason,
the research community has devoted large effort to developing \emph{distance oracles} for approximate distance queries (see~\cite{sommer2014shortest} for a survey).
The key idea is to have an offline pre-processing phase in which an index is built from the network, and then an online querying phase in which the index is used to answer efficiently and approximately the distance queries.

In this paper, we tackle the problem of building distance oracles for \emph{hypergraphs}, which are a generalization of graphs where an edge, called \emph{hyperedge}, represents a $q$-ary relation among vertices.
Hypergraphs are the natural data representation whenever information arises as set-valued or bipartite, i.e., when the entities to be modeled exhibit multi-way relationships beyond simple binary ones: cellular processes~\cite{ritz2015pathway}, protein interaction networks~\cite{feng2021hypergraph}, VLSI \cite{748202}, co-authorship networks~\cite{luo2022toward}, communication and social networks \cite{Tan_Guan_Cai_Qin_Bu_Chen_2014,10.1145/3308558.3313635}, contact networks~\cite{billings2019simplex2vec}, knowledge bases~\cite{fatemi2019knowledge}, e-commerce \cite{10.1145/3219819.3219829}, multimedia \cite{LIU20112255,10.1145/1459359.1459453}, and image processing~\cite{bretto2002hypergraph}.
As taking into account polyadic interactions has been proven essential in many applications, mining and learning on hypergraphs has been gaining considerable research attention recently:
researchers have developed methods for
clustering and classification \cite{7373388,6287378}, representation learning, structure learning and hypergraph neural networks \cite{DBLP:journals/corr/abs-2208-12547,DBLP:conf/aaai/FengYZJG19,DBLP:conf/ijcai/JiangWFCG19}, nodes classification and higher-order link prediction \cite{benson2018simplicial,Zhang_Cui_Jiang_Chen_2018,HyFER,10.1145/2487788.2487802},
% 10.1145/3206025.3206062,},
recommendation \cite{DBLP:conf/kdd/JiFJZT020,10.1145/1873951.1874005,10.1007/s11280-017-0494-5,ZHU2016150},
dense subgraphs and core/truss decomposition \cite{doi:10.1137/16M1096402,luo2022toward,preti2021strud,10.1145/3385416}.
% 10.1145/3437963.3441790,.

Shortest paths in hypergraphs, or \emph{hyperpaths}, have been used in many applications.
\cite{draves2004routing} model a multichannel ad-hoc network as a hypergraph, where each hyperedge represents a group of nodes sharing a channel.
Then, a shortest hyperpath indicates the best route to transmit from source to destination.
\cite{gallo1993directed} represent the bus lines that are of interest for a passenger at a bus stop as a hyperedge, and a shortest hyperpath is a set of attractive origin-destination routes for that passenger.
\cite{krieger2021fast} use a hyperedge to capture a biochemical reaction among reactants, and a hyperpath is a series of reactions that start at receptors and end at transcription factors.
Finally, \cite{ritz2015pathway} find the shortest hyperpaths in a hypergraph modeling protein interactions, to study which proteins and interactions stimulate a specific downstream response in a signaling pathway.
%\cite{aksoy2020hypernetwork,joslyn2020hypernetwork} use hypergraphs for explaining relations among human genes.
%%%%%%%%%%%%%%%%%%%%%%%%%%%%%%%%%%%%%%%%%%%%%%%%%%%%%%%%%%%%%%%%%%%%%%%%%%%%%%

A path is a sequence of adjacent edges and, in graphs, two edges are \emph{adjacent} if they share one vertex.
As hyperedges can contain more than two vertices, a natural extension of the notion of adjacency to hypergraphs is to define two hyperedges adjacent if they share at least $s$ vertices.
The larger the value of $s$, the stronger the relation sought, because the larger is the amount of information (i.e., the vertices) shared between the hyperedges traversed.
This notion of adjacency straightforwardly leads to the definition of $s$\emph{-path}, i.e.,
a sequence of hyperedges such that consecutive hyperedges \revv{share} at least $s$ vertices.
\rev{Previous work~\cite{aksoy2020hypernetwork} shows that $s$-paths and $s$-connected components (i.e., hyperedges connected via $s$-paths) can unveil meaningful, interpretable, and significant structural insight within higher-order data. These insights are often disregarded when the data is modeled as a graph.
For instance, when modeling signaling pathways as ordinary graphs, the representation falls short in capturing critical cellular activities involving the assembly and disassembly of protein complexes, multiway reactions among such complexes, or distinguishing between inactive and active forms of proteins or complexes~\cite{ritz2014signaling}.
Consequently, finding the shortest paths within the underlying graph of the cell-signaling network does not yield the most efficient ways for synthesizing a specific transcription factor from a set of receptors~\cite{klamt2009hypergraphs}, and may even detect non-functional pathways~\cite{de2008can}.}

\begin{figure}[t!]
  \centering
  \includegraphics[width=\linewidth]{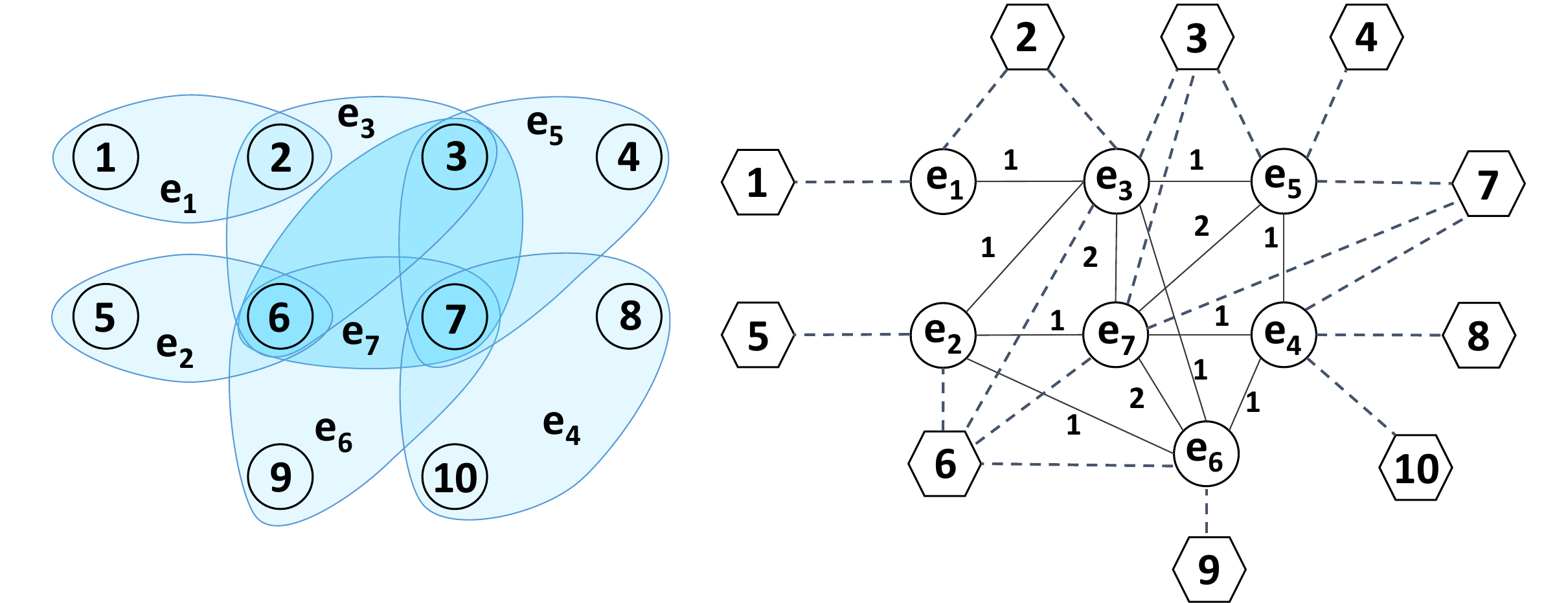}
  \caption{A hypergraph (left) and its line graph (right, see \S\ref{sec:linegraph}) augmented with hexagonal vertices and dashed edges to keep track of  nodes-to-hyperedge membership.}
  \label{fig:toy}
\end{figure}

\begin{example}
Let us assume the hypergraph in Figure~\ref{fig:toy} (left) models contacts among people in an epidemiologic context. For $s=1$, vertex $2$ has the same distance from $7$ and $9$: in fact $2$ belongs to hyperedge $e_3$ which is adjacent to $e_6$ (which contains both $7$ and $9$).
While for $s=2$, vertex $2$ is closer to $7$ (via hyperedge $e_7$ which has two vertices in common with $e_3$) than to $9$, as $e_6$ is no longer adjacent to $e_3$.
As contacts are the cornerstone of the transmission of contagious diseases, if vertex $2$ is infected, the chances that $7$ is infected as well are higher than those of vertex $9$, because $7$ has a stronger connection to $2$.
\end{example}

\spara{Contributions and roadmap.}
We introduce a new type of point-to-point distance queries in undirected hypergraphs,
where a query is defined by the source, the target, and an integer $s$ which indicates the desired level of overlap between consecutive hyperedges in a path.
In particular, we consider variants of this type of query: vertex-to-vertex, hyperedge-to-hyperedge, and vertex-to-hyperedge.
We show that the $s$-distance is a metric only for hyperedge-to-hyperedge queries, while for vertex-to-vertex it is a \emph{semimetric}, as triangle inequality does not necessarily hold when $s > 1$.
Moreover, vertex-to-vertex distance queries can be answered by finding the minimum distance between any two hyperedges that contain the given vertices, and similarly for vertex-to-hyperedge ones.
We thus focus our oracle on hyperedge-to-hyperedge queries.

We show how a simple oracle can be obtained by means of a \emph{line graph}, i.e., a graph representation of a hypergraph.
In it, the hyperedges become vertices, and they have an edge between them if they share vertices in the hypergraph.
However, this approach faces a major limitation in its scalability, as the line graph is typically orders of magnitude larger than the original hypergraph~(\S\ref{sec:linegraph}).
We thus introduce \algo, a framework to build distance oracles with a desired size $Q$, expressed in terms of distance pairs stored by the oracle (\S\ref{sec:hyped}).
Our framework, like many others, is based on the concept of \emph{landmarks}.
In standard graphs, landmarks are vertices for which the oracle stores the distances to all the reachable vertices.
Instead, \algo uses hyperedges as landmarks.
\rev{In contrast to graphs, hypergraphs encompass relations of varying orders, which necessitates the storage of a larger volume of information within the oracle to allow answering $s$-distance queries for different values of $s$.
One of the distinctive challenges in this endeavor is landmark selection. This task is notably more complex than in the case of graphs due to the substantial increase in the number of $s$-connected components as $s$ grows.
Consequently, a well-thought-out selection strategy that keeps into consideration the connectivity structure is essential to ensure that the oracle's size remains within the user-defined limits and that it can approximate as many $s$-distances as possible. Note that this issue does not apply to the graph case, as the oracle is constructed solely for the giant connected component of the input graph.}
% We note that the $s$-connectivity of hypergraphs changes substantially with the increase of the parameter $s$:
% this observation hints at $(i)$ the need to keep into consideration the connectivity structure when selecting landmarks and, $(ii)$ the need of selecting different sets of landmarks for each value of $s$.
We thus devise an algorithm to compute the all the $s$-connected components of a hypergraph up to a max value $\smax$ (\S\ref{sec:ccs}).
Our algorithm combines a \emph{union-find} technique~\cite{tarjan1984worst}, a dynamic inverted index, and a key property of the $s$-distance, to avoid redundant computations.
Then, we devise and compare several different techniques to determine the number of landmarks to assign to each $s$-connected component (\S\ref{sec:lass}), and to select which hyperedges will act as landmarks for each component (\S\ref{sec:la}).
We compare \algo against baselines (\S\ref{subsec:comparison}) and against line-graphs based methods (\S\ref{subsec:comparison2}), to prove its effectiveness in estimating the distances, as well as its efficiency in creating the oracle and answering batches of queries.
Finally, \Cref{sec:casestudy} and \Cref{sec:centrality} prove the usefulness of the $s$-distance oracle in two applications, namely, hypergraph-based recommendation and the approximation of the $s$-closeness centrality of vertices and hyperedges in the context of protein-to-protein interactions.
%\enlargethispage{2\baselineskip}

% !TEX root = ../main.tex
\section{Related Work}\label{sec:related}
\spara{Distance queries in graphs.}
Distance oracles are compact representation of an input graph that enables to answer
point-to-point shortest-path distance queries efficiently~\cite{goldberg2007point,potamias2009fast,sommer2014shortest}, and are assessed according to the space occupied and the query time.
Approximate solutions aim at reducing the space required to store the oracle, with the downside of errors in the answer provided.
\cite{thorup2005approximate} show that, for general graphs, one cannot have both a small oracle and a small error in the answers.
Nonetheless, existing algorithms for general graphs with billions of nodes and edges have proven to be quite accurate and fast~\cite{qi2013toward}.

Even exact solutions may exploit precomputed structures.
\cite{akiba2013fast} propose a distance-aware 2-hop cover set (PLL), further parallelized by~\cite{jin2020parallelizing}.
\cite{xu2016fast} optimize their index for answering batch queries.
While the query time is smaller than that of PLL, the size of the index is much larger.
\cite{li2020scaling} analyze the relation between graph tree width and PLL index size, and proposes a core-tree index (CTL), with the goal of reducing the size requirements of PLL.
The construction of the CTL index is based on a core-tree decomposition of the graph, which results in the identification of a forest of subtrees with bounded tree width and a large core.
Finally, \cite{farhan2018highly} construct a landmark-based index, used to get upper bounds to the distances at query time.
The upper bounds are then used to speed up the computation of the actual distance, which uses a sparsified version of the graph.

\spara{Distance queries in hypergraphs.} Shortest paths in hypergraphs have a long history in the algorithmic literature \cite{gallo1993directed,italiano1989online,nielsen2005finding,ausiello2017directed} and in bioinformatics \cite{ritz2015pathway,franzese2019hypergraph,krieger2021fast}.
However, none of this work considers the problem of approximating shortest-path distance queries by means of an oracle.

\cite{gao2014dynamic} address the shortest path problem for weighted undirected hypergraphs, where paths can only be $1$-paths.
\cite{shun2020practical} focus on single-source shortest paths (SSSP) in weighted undirected hypergraphs and develops a parallel algorithm based on the Bellman-Ford algorithm for SSSP on graphs.
Also in this case, only $1$-paths are considered.

\cite{aksoy2020hypernetwork,joslyn2020hypernetwork} introduce the notion of $s$-walks, where $s$ indicates the overlap size between consecutive hyperedges, and generalize several graph measures to hypergraphs, including
 connected components, centrality, clustering coefficient, and distance.
These generalized measures are then applied to real hypergraphs to prove that they can reveal meaningful structures and properties, which cannot be captured by graph-based methods.
As stated by the authors, the goal of their work is to discuss the need for ad-hoc analytic methods for hypergraphs, whereas the algorithmic aspects have not been taken into account (their analysis is on small hypergraphs up to 22k distinct hyperedges).

\cite{cooley2015evolution} consider higher-order connectivity in random $k$-uniform hypergraphs,  i.e., hypergraphs whose hyperedges have all size $k$.
\cite{lu2011high} consider a slightly different notion of $s$-walks where the hyperedges intersect in exactly $s$ vertices.
Finally, \cite{liu2020parallel,liu2022high} propose parallel algorithms to construct the $s$-line graph of a hypergraph, as well as, an ensemble of $s$-line graphs for various values of $s$.
The experimental evaluation on real-world hypergraphs confirms the hardness of the task, as the algorithm for the ensemble fails on most of the datasets due to memory overflow.

% !TEX root = ../main.tex
\section{Preliminaries}\label{sec:problem}

% \Cref{tab:reference} summarizes the most important notation.
\begin{table}[t!]
\caption{\rev{Notation.}}\label{tab:reference}
\vspace{-3mm}
\small
\begin{center}
\begin{tabular}{rl}
	\toprule
	\rev{\textbf{Symbol}} & \rev{\textbf{Description}} \\
	\midrule
	\rev{$H \doteq (V, E)$}    & \rev{hypergraph} \\
	\rev{$E_i$}                & \rev{hyperedges of size no less than $i$}\\
	\rev{$\mathcal{L}_s$}      & \rev{$s$-line graph of $H$} \\
	\rev{$\mathcal{L}$} 	   & \rev{weighted line graph of $H$} \\
	\rev{$\mathcal{L}_a$}      & \rev{augmented line graph of $H$} \\
	\rev{$\smax$}              & \rev{max value of $s$ considered} \\
	\rev{$\textsf{CC}$}        & \rev{$s$-connected components of $H$ up to $\smax$}\\
    \rev{$\hdist{s}(e,f)$}     & \rev{$s$-distance between $e,f \in E$} \\
    \rev{$\hest{s}(e,f)$}      & \rev{estimated $s$-distance $e,f \in E$} \\
	\rev{$\vdist{s}(u,v)$}     & \rev{$s$-distance between $u,v \in V$} \\
	\rev{$\vest{s}(u, v)$}     & \rev{estimated $s$-distance between $u,v \in V$} \\
	\rev{$\vhdist{s}(u,e)$}    & \rev{$s$-distance between}\\
	                           & \rev{$u \in V$ and $e \in E$} \\
	\rev{$\vhest{s}(u,e)$}     & \rev{estimated $s$-distance between}\\
	                           & \rev{$u\in V$ and $e\in E$} \\
	\rev{$\mathrm{lb}_s$}      & \rev{lower-bound to $\hdist{s}$}\\
	\rev{$\mathrm{ub}_s$}      & \rev{upper-bound to $\hdist{s}$}\\
	\rev{$\mathcal{O}$}        & \rev{distance oracle for $H$}\\
	\rev{$Q$}                  & \rev{max desired oracle size}\\
    \rev{$\dmin$}              & \rev{min size of a connected component}\\
                               & \rev{to be stored in the oracle}\\
    \rev{$\sigma$}             & \rev{landmark selection strategy}\\
	\bottomrule
\end{tabular}
\end{center}
\vspace{-3mm}
\end{table}

We consider an undirected \emph{hypergraph} $H \doteq (V,E)$ where $V = \{v_1,\dots, v_n\}$ is a set of vertices and $E = \{e_1,\dots,e_m\}$ is a set of \emph{hyperedges}.
Each hyperedge $e \in E$ is a set of vertices of cardinality at least $2$, i.e., $e \subseteq V\,\wedge |e|\geq 2$.
The cardinality of a hyperedge is also called \emph{size}, while the \emph{degree} of a vertex $v$ is the number of hyperedges containing $v$.
A \emph{graph} is a hypergraph whose hyperedges have all cardinality equal to $2$.
Finally, given an integer $i$, we denote with $E_i$ the subset of hyperedges with size not lower than $i$.
\rev{\Cref{tab:reference} summarizes the most important notation.}

\revv{
In this paper, we adopt a notion of hyperedge adjacency parameterized by an integer $s$ which indicates the minimum level of overlap between two hyperedges to be considered $s$-adjacent. This notion provides a natural and transparent interpretation of $s$ as an association strength threshold and has been shown to provide meaningful structural insights about hypergraphs~\citep{aksoy2020hypernetwork,joslyn2020hypernetwork}. More importantly for our problem, the $s$-distance between hyperedges defined in this manner is a metric~\citep{aksoy2020hypernetwork}. 
This property enables the construction of accurate landmark-based oracles. In contrast, definitions of hyperedge adjacency based on other set measures such as the Jaccard index lack metric properties, which hinders building accurate hyper-distance oracles. Furthermore, the Jaccard index tends to undervalue relationships between large and small hyperedges (even when one is contained in the other), and emphasizes relations among smaller hyperedges, potentially skewing the overall assessment of association strength.}

\begin{definition}[$s$-adjacent, $s$-walk, $s$-path~\cite{aksoy2020hypernetwork}]
\rev{Given $s \in \mathbb{N}^+$, a $s$-walk of length $l \geq 2$ is a sequence of $l$ hyperedges $[e_1, \dots, e_l]$ such that, for any $j \in [1, l-1]$, $e_j$  and $e_{j+1}$
are $s$-adjacent, that is,  $\lvert e_{j} \cap e_{j+1} \rvert \geq s$.
An $s$-path is a $s$-walk where for each $i \neq j \in [1, l]$ it holds that $e_i \neq e_j$.}
Two hyperedges $e$ and $f$ are $s$-connected iff there exists an $s$-path $[e, \dots, f]$.
%\rev{When $l = 2$, i.e., the $s$-path is $[e, f]$, we say that $e$ and $f$ are $s$-adjacent.}
\end{definition}
\rev{A $1$-walk in the hypergraph in \Cref{fig:toy} is $[e_1, e_3, e_5, e_6, e_3]$; a $2$-walk is $[e_3, e_7, e_5, e_7, e_6]$; a $1$-path is $[e_2, e_6, e_4]$; and a $2$-path is $[e_3, e_7, e_6]$. Since the max overlap size between any two hyperedges is $2$, the hypergraph is $3$-disconnected.}
\rev{The pair of hyperedges $[e_3, e_7]$ is an example of $2$-adjacent hyperedges, whereas $[e_1, e_3]$ are only $1$-adjacent. We can easily see that a pair of $s$-adjacent hyperedges are also $(s-1)$-adjacent.}

Clearly, $s$-paths generalize graph paths to hypergraphs. In fact, a path in a graph is equivalent to a $1$-path in the hypergraph representation of the graph, because any pair of adjacent edges in a graph path shares exactly $1$ vertex.
The existence of $s$-paths between hyperedges defines equivalence classes of hyperedges:
\begin{definition}[$s$-connected components~\rev{\cite{aksoy2020hypernetwork}}]
Given a hypergraph $H$, a set of hyperedges $C \subseteq E$ is an \emph{s-connected component} iff \textbf{(i)} there exists an $s$-path between any pair of hyperedges $e,f \in C$, and \textbf{(ii)} $C$ is maximal, i.e., there exists no $J \supset C$ that satisfies the first condition.
We say that $H$ is \emph{s-connected} iff there exists a single $s$-connected component, i.e., $C = E$.
\end{definition}

The number of hyperedges in $C$ is also known as the \emph{size} of $C$.
The \emph{density} of $C$ is measured as the ratio between the number of $s$-adjacent hyperedges in $C$ and the number of possible pairs of hyperedges in $C$, which is equal to $\binom{|C|}{2}$.

The \emph{s-distance} between two hyperedges $e$ and $f$, which we denote as $\hdist{s}(e, f)$, is defined as the length of the shortest $s$-path between $e$ and $f$, decreased by 1.\footnote{The $s$-distance is often defined as the length of the shortest $s$-path. \revv{We subtract 1 to align with graph theory conventions}, where the distance between edges is the number of vertices in a shortest path between them, \revv{making} adjacent edges connected by a path of length 2 but at distance 1.}
That is, the distance between two $s$-adjacent hyperedges is 1. Moreover, $\hdist{s}(e, e) = 0$.
If no $s$-path exists between $e$ and $f$ (i.e., they belong to different s-connected components), then $\hdist{s}(e,f)=\infty$. If $|e| < s$, then $\hdist{s}(e,f)=\infty$ for all $f \neq e \in E$.

\begin{definition}[vertex-to-vertex $s$-distance]
The \emph{s-distance} between two vertices $u,v \in V$ is
\rev{
\begin{align}
\vdist{s}(u,v) \doteq	
	\begin{cases}
	0 & \text{ if } u = v \,, \\
	\min\limits_{e_1 \ni u, e_2 \ni v}\hdist{s}(e_1, e_2) + 1 & \text{ otherwise.}
	\end{cases}
\end{align}
}
\label{def:s-distance}
\end{definition}
%Similarly, we can define a vertex-to-hyperedge distance query:

\begin{definition}[vertex-to-hyperedge $s$-distance]
\rev{The \emph{s-distance} between a vertex $u \in V$ and a hyperedge $e \in E$ is
$\quad \vhdist{s}(u,e) \doteq \min\limits_{e_1 \ni u}\hdist{s}{e_1, e}$.
}
\end{definition}

While the $s$-distance between hyperedges is a metric~\cite{aksoy2020hypernetwork}, the $s$-distance between two vertices, or between a vertex and an hyperedge, is a \emph{semimetric}, i.e., the triangle inequality does not necessarily hold when $s > 1$.
As an example, let consider the hypergraph in Figure~\ref{fig:toy} (left).
We can see that $\vdist{2}(1,4)=\infty$, because there is no $2$-path between hyperedges containing $1$ and $4$.
In fact, only $e_1$ contains vertex $1$, and the maximum overlap between $e_1$ and any other hyperedge is $1$.
In contrast, $\vdist{2}(2,4)=3$, thanks to the $2$-path $[e_3, e_7, e_5]$; and $\vdist{2}(1,2)=1$, thanks to the common hyperedge $e_1$.
Since $\vdist{2}(1,4) > \vdist{2}(1,2)+\vdist{2}(2,4)$, the measure $\vdist{s}$ does not satisfy the triangle inequality.\footnote{\revv{An alternative definition of vertex-to-vertex $s$-distance that is a metric could be considered. Let the dual hypergraph be the one obtained by swapping the roles of vertices and hyperedges: hyperedges become vertices, and each vertex in the original hypergraph becomes a hyperedge that connects all the vertices in the dual that correspond to the hyperedges of the original hypergraph by which it was contained. We can compute the hyperedge-to-hyperedge $s$-distance in this dual hypergraph and obtain a metric vertex-to-vertex $s$-distance in the original hypergraph.}

\revv{However, note that the vertex-to-vertex $s$-distance as in \Cref{def:s-distance} and the hyperedge-to-hyperedge $s$-distance in the dual hypergraph yield distinct results.
For instance, $s$-connected vertices in the hypergraph may be at infinite distance in the dual.
In fact, an $s$-path in the hypergraph is a sequence of hyperedges such that consecutive hyperedges share at least $s$ common vertices, whereas an $s$-path in the dual is a sequence of vertices such that consecutive vertices belong to at least $s$ common hyperedges.}

\revv{Both definitions are valid and could be adopted depending on the applications at hand. Our framework can also handle this alternative definition of vertex-to-vertex $s$-distance, by simply applying it to the dual hypergraph. Of course, this leads to a separate oracle which could not be used to answer hyperedge-to-hyperedge $s$-distance queries in the original hypergraph.
Therefore, henceforth we focus on the vertex-to-vertex $s$-distance as per \Cref{def:s-distance}.
This allows to answer all three types of queries with a single oracle.}}

%The vertex-to-vertex $s$-distance plays a crucial role in measuring the proximity between vertices in a hypergraph, accounting for the strength of hyperedge relations.
%A higher $s$ signifies greater similarity between consecutive hyperedges, reflecting increased relatedness among $s$-connected vertices.

Given that hyperedge-to-hyperedge distance is the basis for computing also the other two types of queries,
we focus our attention on building oracles for this type of distance query, i.e.,
given $e,f \in E$ and $s\in \mathbb{N}^+$, return, as fast as possible, $\hdist{s}(e,f)$ or a good approximation $\tilde{\hdist{s}}(e,f) = \hdist{s}(e,f) + \epsilon$ with $\epsilon \approx 0$.

In many applications, the best value of $s$ might not be specified; on the contrary, the user might be interested in understanding how close two entities are for different values of $s$.
For instance, in recommender systems
% (see \Cref{subsec:casestudy})
one may want to rank the results of the queries trading-off between distance and overlap size: a close
entity for a small $s$ may be as relevant as an entity which is more distant, but for a larger $s$.
In these cases, especially if the connectivity and the dimension of the hypergraph are not known apriori, the user needs to ask several queries with different values of the parameter $s$.
For this reason, we study a more general problem, i.e., to compute the \emph{distance profile} of $e$ and $f$:
\begin{problem}[distance profiling]\label{prob:dp}
Given $e \neq f \in E$, compute $\hdist{s}(e,f)$ for all $s \leq \min(|e|, |f|)$.
\end{problem}

The stopping threshold for the distance profiles derives from the fact that for all $s > \min\{|e|,|f|\}$, $\hdist{s}(e,f)=\infty$.

This problem can be solved at query time, by executing a bidirectional BFS for each value of $s$.
This strategy, however, leads to large query times; hence it is not viable for real-time query-answering systems.
A common approach for fast query-answering is to pre-compute all the distance pairs and store them in a distance table.
This approach, however, requires to store a huge amount of information, and hence is characterized by high space complexity.
A trade-off between time and space is represented by the so-called \emph{distance oracle}.
A distance oracle is a data structure that stores less information than a distance table, but allows a query to be answered faster than executing a BFS.
Distance oracles can be designed to answer queries either exactly or approximately.
In line with the literature on shortest path queries in graphs, we opt for approximate in-memory solutions, to further reduce the space complexity~\cite{gubichev2010fast}.

% !TEX root = ../main.tex
\section{Line-graph based oracle}\label{sec:linegraph}
One of the tools widely used in hypergraph analysis is the \emph{line graph}, which represents hyperedges as vertices and intersections between hyperedges as edges:

\begin{definition}[line graph~\cite{berge1984hypergraphs}]
Let $H = (V,E)$ be a hypergraph. The \emph{line graph} of $H$, denoted as $\mathcal{L}(H)$, is the weighted graph on vertex set $E$ and edge set $\{(e, f) \subseteq E \times E \,|\, e \neq f \,\wedge\,|e \cap f| \geq 1\}$.
Each edge $(e,f)$ has weight $\omega(e,f) = |e \cap f|$.
\end{definition}

The $s$\emph{-line graph} of $H$, denoted $\mathcal{L}_s(H)$, is its line graph from which only edges with weight at least $s$ are retained.
It is straightforward to see that the $s$-distance $\hdist{s}(e, f)$ between two hyperedges $e$ and $f$ in the hypergraph $H$, corresponds to the distance between their corresponding vertices in the $s$\emph{-line graph} $\mathcal{L}_s(H)$.

\begin{example}
Figure \ref{fig:toy} shows an hypergraph (left) and its corresponding line graph (right) augmented with a vertex for each vertex in the hypergraph (denoted with hexagons) and an edge for each vertex-to-hyperedge membership (denoted with dashed lines).
We can easily verify that the distance between $e_3$ and $e_5$ in the $2$-line graph is $2$ (they are connected via the shortest path $[(e_3, e_7), (e_7, e_5)]$ of edges with weight $\geq 2$), while $\hdist{2}(e_3, e_5) = 2$ (via the 2-path $[e_3, e_7, e_5]$).
\end{example}

This observation hints a way to produce a distance oracle for a given hypergraph $H$ based on its line graph $\mathcal{L}(H)$. At query time, given hyperedges $e$ and $f$ and the desired value of $s$, one can obtain $\mathcal{L}_s(H)$ from $\mathcal{L}(H)$, retrieve the vertices corresponding to $e$ and $f$ and compute their distance in $\mathcal{L}_s(H)$. As the line graph is a standard graph, one can preprocess it to build a distance oracle using any state-of-the-art method for graphs (see experiments in Section \ref{subsec:comparison}).

\spara{Limitations.} The main limitation of such an approach is that line graphs are typically much larger than the original hypergraphs, as their number of edges is in the order of \rev{$O(\mathsf{d}_{\mathsf{MAX}}^2)$, where $\mathsf{d}_{\mathsf{MAX}}$ is the max number of $s$-adjacent hyperedges of a hyperedge in $H$}.
The complexity worsens when we need to answer also vertex-to-vertex $s$-distance queries:
in this case, the line graph needs to be augmented with a new vertex $u_v$ for each $v \in V$, and a new edge between $u_v$ and each vertex in the line graph that corresponds to a hyperedge including $v$ (see Figure \ref{fig:toy} (right)). As a consequence, the \emph{augmented line graph} has $|V| + |E|$ vertices of different types (those denoted with circles and those with hexagons in \Cref{fig:toy}), and up to $|V||E| + \rev{\mathsf{d}_{\mathsf{MAX}}^2}$ edges.

\begin{table}[!t]
\centering
\caption{\rev{Number of nodes and hyperedges in the hypergraph $H =(V,E)$, number of edges in the corresponding line graph $\mathcal{L}$, number of nodes and number of edges in the corresponding augmented line graph $\mathcal{L}_a$, disk size of $H$ ($\textsf{DS}_H$), and disk size of $\mathcal{L}$ ($\textsf{DS}_\mathcal{L}$).}}
\label{tbl:linegraphs}
\resizebox{\columnwidth}{!}{
\begin{tabular}{lrrrrrrr}
	\toprule
	\textbf{Dataset} & $\mathbf{|V|}$ & $\mathbf{|E|}$ & $\mathbf{|E_\mathcal{L}|}$ & $\mathbf{|V_{\mathcal{L}_a}|}$ & $\mathbf{|E_{\mathcal{L}_a}|}$ & \rev{$\textsf{DS}_H$} & \rev{$\textsf{DS}_\mathcal{L}$}\\
	\midrule
	\textbf{NDC-C} & \num{1.1}K & \num{1}K & \num{36}K & \num{2.2}K & \num{42}K & \rev{\num{28}K} & \rev{\num{1.2}M}\\
	\textbf{ETFs} & \num{2.3}K & \num{2}K & \num{47}K & \num{4.5}K & \num{53}K  & \rev{\num{28}K} & \rev{\num{444}K}\\
	\textbf{High} & \num{327} & \num{8}K & \num{593}K & \num{8.1}K & \num{611}K & \rev{\num{72}K} & \rev{\num{5.8}M}\\
	\textbf{Zebra} & \num{10}K & \num{10}K & \num{104}K & \num{20}K & \num{154}K & \rev{\num{352}K} & \rev{\num{3.3}M}\\
	\textbf{NDC-S} & \num{5.3}K & \num{10}K & \num{2.4}M & \num{15}K & \num{2.4}M & \rev{\num{256}K} & \rev{\num{33}M}\\
	\textbf{Primary} & \num{242} & \num{13}K & \num{2.2}M & \num{13}K & \num{2.3}M & \rev{\num{152}K} & \rev{\num{23}M}\\
	\textbf{Enron} & \num{50}K & \num{106}K & \num{153}M & \num{156}K & \num{153}M & \rev{\num{1.6}M} & \rev{\num{1.9}G}\\
	\textbf{Epinions} & \num{756}K & \num{107}K & \num{32}M & \num{863}K & \num{46}M & \rev{\num{90}M} & \rev{\num{944}M}\\
	\textbf{DBLP} & \num{1.9}M & \num{2.4}M & \num{125}M & \num{4.4}M & \num{133}M & \rev{\num{51}M} & \rev{\num{2}G}\\
    \textbf{Threads-SO} & \num{2.7}M & \num{7}M & \num{263}M & \num{9.6}M & \num{277}M & \rev{\num{99}M} & \rev{\num{3.9}G}\\
	\bottomrule
\end{tabular}
}
\vspace{-3mm}
\end{table}

\Cref{tbl:linegraphs} reports several statistics for $8$ real-world hypergraphs: number of vertices, number of hyperedges, number of vertices and edges in the line graph, number of vertices and edges in the augmented line graph, \rev{disk usage of the hypergraph, and disk usage of the line graph}.
On average, the line graph has 238x more edges than the related hypergraph has hyperedges, with Enron being the hypergraph with the largest increase (1435x).
Similarly, the augmented line graph has 254x more edges, on average. %, with Enron being the hypergraph with the largest increase (1438x).
Consequently, a distance oracle built on top of a line graph may require significantly more space. A \rev{second} issue with a line-graph-based solution is that creating the line graph is a challenge in itself.
Few approaches have been proposed in the literature, which, however, aim at generating the $s$-line graph for a given $s$~\cite{liu2020parallel}, and hence require a considerable amount of redundant work to obtain all the $s$-line graphs up to a max $\smax$.
\rev{A third issue is that} the state-of-the-art distance oracles for graphs are not designed to consider two different types of vertices, as it is the case with the augmented line graph needed to answer
vertex-to-vertex $s$-distance queries.
Thus, their adoption would not be straightforward in this case (in fact, in \Cref{subsec:comparison} we experiment with the line-graph-based approach only for hyperedge-to-hyperedge queries).
\rev{Finally, the fidelity of line graphs in accurately representing hypergraphs remains uncertain, as demonstrated by Kirkland~\cite{kirkland2018two}: instances exist where non-isomorphic hypergraphs yield identical weighted line graphs. Consequently, oracles built on top of line graphs might result in less accurate distance approximations, especially for specific types of hypergraphs.} 
% !TEX root = ../main.tex
\section{Landmark-based oracle}\label{sec:solution}
In light of the limitations discussed above, we propose an approach based on the idea of \emph{landmarks}~\cite{potamias2009fast} that allows building oracles of a predefined size that operate directly on the hypergraph, thus avoiding going through its line graph.

In the standard graph setting, given a set of \emph{landmark} vertices $L \subset V$, a distance oracle is built by storing the exact distances $d(l,u)$ from each landmark $l \in L$ to each vertex $u \in V$.
Therefore, a larger number of landmarks implies a higher space complexity.
Then, the approximate distance $\hat{d}(u,v)$ between $u$ and $v$ is given by $\hat{d}(u,v) = \min_{l \in L}{d(l,u)+d(l,v)}$.
Since the distance between graph vertices is a metric, the approximate distance satisfies $\hat{d}(u,v) \geq d(u,v)$, and hence constitutes an upper bound to the real distance.
In addition, a lower bound to $d(u,v)$ can be computed as $\max_{l \in L}\left|d(l,u) - d(l,v)\right|$.

\begin{figure}[t!]
    \centering
    \includegraphics[width=\columnwidth]{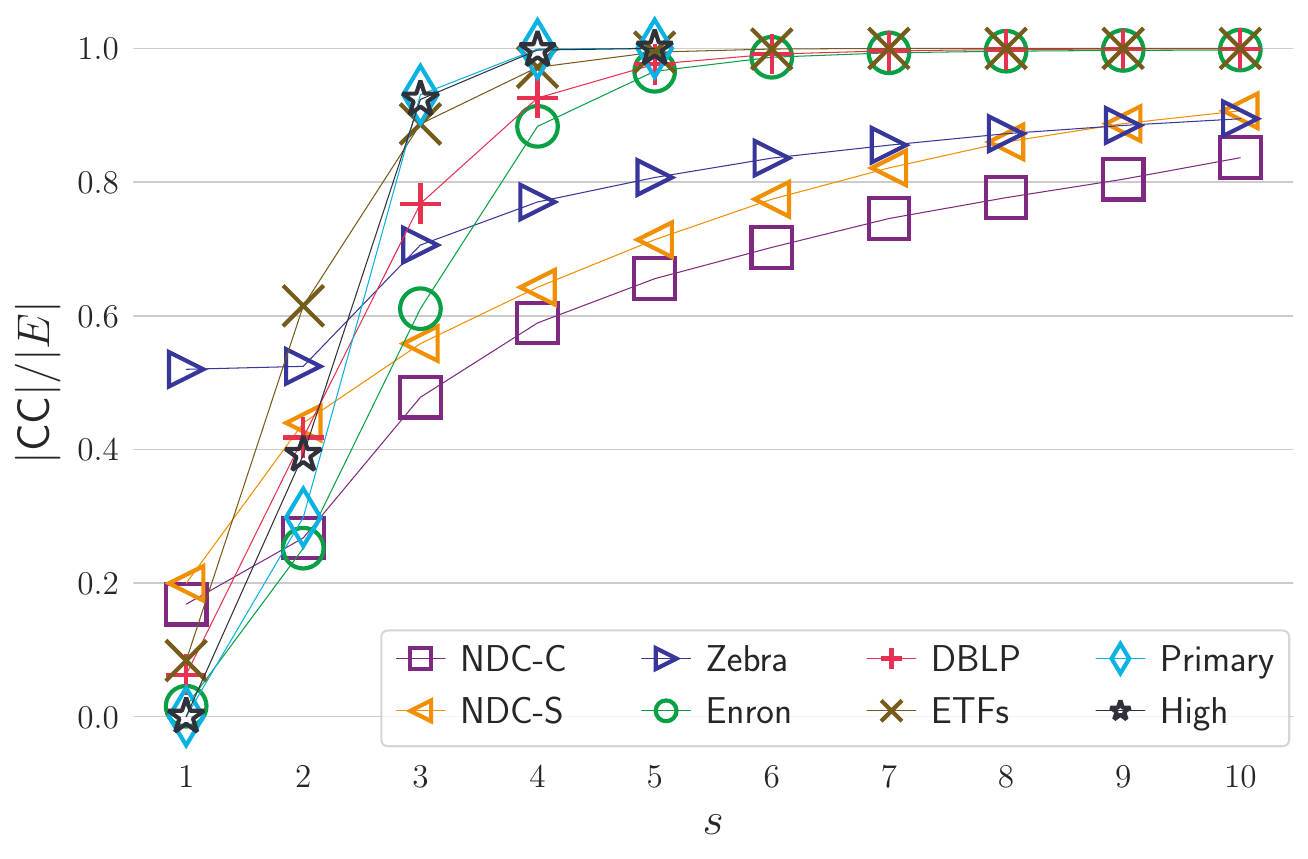}
    \caption{Number of $s$-connected components $|\textsf{CC}|$ (normalized to  the number of hyperedges $|E|$) as a function of $s$. Lower values indicate higher connectivity.}
    \label{fig:ccs}
\end{figure}

In our setting, one first key idea is to use hyperedges, instead of vertices, as landmarks. Recall in fact that, in hypergraphs, only the hyperedge-to-hyperedge distance satisfies the triangle inequality, a crucial property for landmark-based solutions.
The second key idea is to select different sets of landmarks for different values of the parameter $s$.
\Cref{fig:ccs} shows how the connectivity of several real-world hypergraphs changes as the value of the parameter $s$ increases.
For $s=1$, the hypergraphs are almost connected, but as $s$ increases, the hyperedges form smaller $s$-connected components.
As a consequence, a set of landmarks selected for a certain $s$ may not be effective enough for approximating the $(s+1)$-distances.
This fact motivates the need for different sets of landmarks.

Next, we introduce our framework, \algo (\underline{Hyp}ergraph \underline{E}stimation of $s$-\underline{D}istances), which exploits a connectivity-based landmark assignment strategy and the topology of the $s$-connected components of the hypergraph, to construct a landmark-based distance oracle with predefined size.
Finally, we discuss its complexity in \Cref{sec:complexity}.

\subsection{The \algo Framework}\label{sec:hyped}
\algo (\Cref{alg:create_oracle}) bases its landmark selection on the $s$-connected components $\textsf{CC}$ of $H$,
given an integer $\smax$, the maximum value of the parameter $s$ for which to estimate the $s$-distances.
\begin{algorithm}[!t]
\footnotesize
\begin{algorithmic}[1]
\Require Hypergraph $H$, Max $s$ $\smax$
\Require Min Dim $\dmin$, Max Oracle Size $Q$
\Require Importance Factors $\Omega$
\Require Landmark Selection Strategy $\sigma$
\Ensure Distance Oracle $\mathcal{O}$
\State $\mathcal{O}.\textsf{CC} \gets \Call{findConnectedComponents}{H, \smax}$ %\Comment{\textbf{memberships}}
\State $\mathcal{O}.\textsc{avgD} \gets \varnothing$
\For{$i \in [2, \dmin]$} $\mathcal{O}.\textsc{avgD}[i] \gets \Call{approxAvgDist}{i}$ \label{line:avg}
\EndFor
\State $L \gets \Call{selectLandmarks}{\mathcal{O}.\textsf{CC}, Q, \dmin, \Omega, \sigma}$
\For{$s \in [1, \smax]$} $\mathcal{O}[s] \gets \Call{populateOracle}{H, L[s], s}$  \label{line:lab}
\EndFor
\State \Return $\mathcal{O}$
\end{algorithmic}
\caption{\textsc{HypED}}\label{alg:create_oracle}
\end{algorithm}

\begin{algorithm}[!t]
\footnotesize
\begin{algorithmic}[1]
    \Require HyperGraph $H$, Max $s$ $\smax$
    \Ensure Connected Components $\textsf{CC}$
    \State $I \gets \varnothing$;$\textsc{OP} \gets \varnothing$; $\textsc{CP} \gets \varnothing$
    \For{$s \in [\smax, 1]$}
        \State $E_s \gets e \in E \text{ s.t. } |e| \geq s$
        \State $\tilde{\textsf{CC}} \gets [\{e\} \text{ for each } e \in E_s]$
        \State initialize $\tilde{\textsf{CC}}$ using $\textsf{CC}_{s+1}$\label{line:init}
        \For{$e \in E_s$ not indexed yet}
            \For{$v \in e$}
                $I[v] \gets I[v] \cup \{e\}$\label{line:index}
            \EndFor
        \EndFor
        \State $\textsc{OV} \gets \varnothing$
        \For{$v \in I$}
            \For{$(e_1, e_2) \in I[v]$}
                \If{$e_1$ and $e_2$ are not in the same $c \in \tilde{\textsf{CC}}$}
                    \State $\textsc{OV}[(e_1, e_2)] \gets \textsc{OV}[(e_1, e_2)] + 1$\label{line:update}
                    \If{$\textsc{OV}[(e_1, e_2)] \geq s$}
                        \State $\textsc{OP} \gets \textsc{OP} \cup \{(e_1, e_2, \textsc{OV}[(e_1, e_2)])\}$
                        \State $\tilde{\textsf{CC}} \gets \Call{unionFind}{\tilde{\textsf{CC}}, e_1, e_2}$
                    \EndIf
                \Else
                    $\:\: \textsc{CP} \gets \textsc{CP} \cup \{(e_1, e_2)\}$
                \EndIf
            \EndFor
        \EndFor
        \State $\textsf{CC}[s] \gets \tilde{\textsf{CC}}$
    \EndFor
    \State \Return $\textsf{CC}$
\end{algorithmic}
\caption{\textsc{findConnectedComponents}}\label{alg:ccs}
\end{algorithm}
\begin{table}[!t]
    \centering
    \scriptsize
    \caption{Approximate distances in topologies with different numbers of elements $n$ and connections $m$.}
    \label{tbl:distances}
    \vspace{-3mm}
    \begin{tabular}{lrrrrrrrr}
      & \multicolumn{7}{c}{$\mathbf{m}$}\\
      \toprule
        \multicolumn{1}{l|}{$\mathbf{n}$} & $n-1$ & $n$ & $n+1$ & $n+2$ & $n+3$ & $n+4$ & $n+5$\\
        \midrule
        2 & 1 & & & & & & \\
        3 & 1.3 & 1 & & & & &\\
        4 & 1.55 & 1.3 & 1.16 & 1 & & &\\
        5 & 1.8 & 1.58 & 1.42 & 1.3 & 1.2 & 1.1 & 1\\
      \bottomrule
    \end{tabular}
    \vspace{-3mm}
\end{table}

\spara{Offline phase: oracle construction.} The first step is the computation of $\textsf{CC}$ (\Cref{alg:ccs}).
Then, the oracle stores, for each $e \in E$ and each $s \leq \min([\smax, |e|])$, the id of the $s$-connected component that includes $e$.
At query time, this information speeds up the distance estimation for hyperedges in different components.

For small $s$-connected components, the $s$-distance between their hyperedges can be bounded by a small constant.
Therefore, it can be approximated accurately without using landmarks, by exploiting their topology (e.g., wedge, triangle, star).
Knowing the topology of a component $c$ allows us to exactly compute the average distance between its hyperedges.
However, the identification of the topology of $c$ is a hard problem known as \emph{graph isomorphism}.
To reduce the time and space complexity of our approach, we estimate the approximate distance between hyperedges in $c$ as the expected average distance between pairs of hyperedges across topologies with the same size of $c$, and store a value for each size $i$ up to a max integer $\dmin$ (\Cref{alg:create_oracle} \Cref{line:avg}).
For example, for $|c|=3$ there are two topologies: the wedge and the triangle.
In the former, the average distance between pairs of nodes is $1.33$, while in the latter, it is 1. Therefore, the approximate distance is $1.16$.
This computation is doable for topologies with up to $5$ hyperedges, as they can be easily enumerated, and their pairwise distances can be easily listed.
The approximate distances for topologies with $n$ elements and $m$ connections is reported in \Cref{tbl:distances}.
We observe that in the online phase, if the algorithm has access to the input hypergraph, distances between hyperedges in small $s$-connected components can be efficiently computed using bidirectional BFSs. However, in our case, we assume that during query time, we only have access to the distance oracle.

For each $s$-connected component of size larger than $\dmin$, the subroutine \textsc{selectLandmarks}
 assigns a budget of landmarks and selects them.
In particular, \algo takes as input a desired oracle size $Q$ indicating the maximum number of distance pairs that we are willing to store in the oracle.
This threshold is used
to select the appropriate amount of landmarks in the $s$-connected components.
Finally, given the sets of landmarks $L_s$ for each $s \leq \smax$, the oracle finds and stores the exact $s$-distance between each $l \in L_s$ and each $e \in E_s$ reachable from $l$ (\Cref{alg:create_oracle} \Cref{line:lab}).

\spara{Online phase: estimating the distance profile. }%\label{sec:online}
 Given the sets of landmarks $L_s$ selected for each $s$ of interest, \algo computes the lower ($\mathrm{lb}_s$) and upper ($\mathrm{ub}_s$) bounds to the $s$-distance between any pair of hyperedges $e$ and $f$ as follows:
\begin{align}
\mathrm{lb}_s(e, f) = \max_{l \in L_s}\left|\hdist{s}(l,e) - \hdist{s}(l,f)\right| \label{eq:lb} \\
\mathrm{ub}_s(e, f) = \min_{l \in L_s}(\hdist{s}(l,e) + \hdist{s}(l,f)) \label{eq:ub}
\end{align}

Then, it generates the distance profile of $e$ and $f$, by computing, for each $s \in [1, \smax^*]$ with $\smax^* = \min(\smax, |e|, |f|)$, the approximate $s$-distance $\hest{s}(e, f)$ as:
\begin{small}
$$
\hest{s}(e, f) = 	\begin{cases}
						\infty & \text{ if } e \in c^s_i \wedge f \in c^s_j \wedge i \neq j ,\\[5pt]
						\textsc{avgD}(|c^s|) & \text{ if } e,f \in c^s \wedge |c^s| \leq \dmin ,\\[5pt]
						\frac{\left( \mathrm{lb}_s(e,f) + \mathrm{ub}_s(e,f) \right)}{2} & \text{ otherwise.}
					\end{cases}
$$
\end{small}
where \textsc{avgD} is the average pairwise distance in a graph with $|c^s|$ vertices, computed in \Cref{alg:create_oracle} \Cref{line:avg}.

Given two query vertices $u$ and $v$, the distance profile of $u$ and $v$ is obtained by calculating the approximate $s$-distance $\vest{s}(u,v)$, for each $s \in [1, \smax^*]$ with $\smax^* = \min(\smax, |e^u|, |e^v|)$, with $e^u$ ($e^v$) being the largest hyperedge including $u$ ($v$), as:
$$
\vest{s}(u,v) = \min\limits_{e \ni u, f \ni v}\hest{s}(e, f).
$$

The approximate $s$-distances are further refined based on the following observation:
\begin{observation}\label{obs1}
For any $e, f \in E$, $$\hdist{s}(e,f) \leq \hdist{s+1}(e,f).$$
\end{observation}
In particular, the lower-bound $\mathrm{lb}_s(e,f)$ for $s \in [2, \smax^*]$ can be improved as $\mathrm{lb}_s(e,f) = \max{(\mathrm{lb}_s(e,f), \mathrm{lb}_{s-1}(e,f))}$;
while the upper-bound $\mathrm{ub}_s(e,f)$ for $s \in [1, \smax^*-1]$ can be improved as $\mathrm{ub}_s(e,f) = \min{(\mathrm{ub}_s(e,f), \mathrm{ub}_{s+1}(e,f))}$.

\subsection{Finding the s-Connected Components}\label{sec:ccs}
The $s$-connected components of the hypergraph are a key factor for the selection of landmarks.
Their computation is the bottleneck of the offline phase of our algorithm.
For this task, we propose a technique that combines the \emph{union-find} approach~\cite{tarjan1984worst} and a dynamic inverted index, while exploiting \Cref{obs1}.
We first describe how union-find works, and then explain our technique, which is outlined in \Cref{alg:ccs}.

\spara{Union-find}
algorithms have been successfully applied to find connected components in graphs~\cite{manne2009scalable}.
Here we describe a more advanced version with two types of optimizations.

Given a set of vertices $V$ and a set of edges $E$, it iterates over the edges to find disjoint sets of connected vertices, i.e., groups of vertices connected via paths, but isolated from each other.
The underlying data structure is a forest, where each tree represents a connected set, and each vertex is associated with \emph{(i)} a \emph{rank} initially set to $0$ and \emph{(ii)} a pointer to the parent.
Each vertex starts as a tree by itself and hence points to itself.
Then, the connected components are found via the application of two operators: \emph{union-by-rank} and \emph{find} enhanced with \emph{path compression}.
Given an edge $(u,v)$, \emph{union-by-rank} merges the tree of $u$ and that of $v$, based on the rank associated to their root: if the rank is different, the root with lowest rank is set to point to the other root; if the two ranks are equal, one of the roots is set to point to the other one, and the rank of the latter is increased by 1.
Given a vertex $u$, \emph{find} returns the root of the tree of $u$, and, for all the vertices traversed, sets their pointer to the root.
This way, any subsequent \emph{find} for these vertices will be faster.

\spara{\Cref{alg:ccs}.}
According to Observation~\ref{obs1}, if two hyperedges are in the same $s$-connected component, they also belong to the same $(s-1)$-connected component.
Therefore, we search for the $s$-connected components in decreasing order of $s$, so that the $s$-connected components can be used to initialize the $(s-1)$-connected components (\cref{line:init}).
The dynamic inverted index $I$ stores, for each vertex $v$, the hyperedges including $v$.
This index is updated at each iteration $s$, including the hyperedges $e$ not yet considered, i.e., those with size $s$ (\cref{line:index}).
Thanks to $I$, we can examine only pairs of hyperedges that actually overlap. The partial overlap is stored in a dictionary $\textsc{OV}$.
Each time a pair $(e_1, e_2)$ is considered, we first check if we already know that they belong to the same connected component.
This can happen in two cases: \emph{(i)} $e_1$ and $e_2$ belong to the same $(s+1)$-connected component, or \emph{(ii)} union-find merged their trees in a previous call.
If this is not the case, we update their partial overlap (\cref{line:update}).
If the overlap exceeds the min overlap size $s$, then $e_1$ and $e_2$ are $s$-connected, and hence we can call \textsc{unionFind} to propagate this information and update $\tilde{\textsf{CC}} $.
During the search of the $s$-connected components, we discover partial overlaps $\textsc{OV}[(e_1, e_2)]$ between pairs of hyperedges.
In addition, thanks to the transitive property, we may find that two hyperedges belong to the same $s$-connected component, without computing their overlap.
We store these pairs in a set $\textsc{CP}$, and the partial overlaps in a set $\textsc{OP}$.
These sets are exploited in the initialization of the sets of $s$-adjacent hyperedges, which are needed to find the $s$-distances between hyperedges.
In particular, if the partial overlap between two hyperedges is $\textsc{OV}[(e_1, e_2)] = o$, then, $e_1$ and $e_2$ are $s$-adjacent for each $s \leq o$.
Conversely, for each pair of hyperedges in $\textsc{CP}$, we need to compute their overlap, as we only know that they are $s$-connected but not whether they are $s$-adjacent. To save space, we do not explicitly store the $s$-connected components of hyperedges with size lower than $s$, as such hyperedges are s-disconnected from the rest of the hypergraph.
The oracle can still compute the vertex-to-vertex $s$-distance between vertices in the same hyperedge of size lower than $s$, thanks to a vertex-to-hyperedge map created at query time.
This map stores, for each vertex, the ids of the hyperedges containing that vertex.
This way, we can determine the 1-hop neighbors of each vertex.

%\note{Secondo me questo è un dettaglio, un corner case... non starei a scrivere sta roba }

Next, we show how \algo exploits the $s$-connected components to guide the landmarks assignment and selection process.
It explicitly divides the process into two phases: a landmark assignment (LA) phase, where the goal is to find the number of landmarks $l_c$ to assign to each $s$-connected components $c$, for each $s \in [1, \smax]$; and a landmark selection (LS) phase, where the goal is to select, for each component $c$, $l_c$ hyperedges as landmarks.
As discussed previously, only the connected components with size greater than the lower bound $\dmin$ are taken into consideration.

\subsection{Landmark Assignment}\label{sec:lass}
We propose two different strategies for the LA phase: a sampling-based strategy and a ranking-based one.
Both strategies require the set $\textsf{CC} = \left\{\textsf{CC}_1, \ldots, \textsf{CC}_{\smax}\right\}$ with $\textsf{CC}_i = \left\{c^i_1, \ldots, c^i_{|\textsf{CC}_i|}\right\}$ of $s$-connected components up to a maximum value $\smax$, the target oracle size $Q$, and a set of importance factors $\Omega$.
These strategies assign a number of landmarks to each $s$-connected component, such that the total size of the oracle is bounded by $Q$.
Once the assignment is found, the selection of the landmarks within each component (LS phase) is executed according to a strategy $\sigma$ (\Cref{sec:la}).

The following observation tells us that the assignment strategy should prioritize larger connected components:
\begin{observation}\label{obs3}
The larger a connected component $c$, the higher the probability that a distance query involves two hyperedges in $c$.
\end{observation}

Furthermore, the following observation suggests that connected components with more vertices should be preferred:
\begin{observation}\label{obs5}
Given equal number of hyperedges, the more the vertices in a connected component $c$, the sparser is $c$, the higher the probability that each hyperedge covers a lower number of shortest paths.
\end{observation}
In fact, the lower the number of vertices in the connected component $c$, the more the hyperedges in $c$ overlap with each other, and therefore, the shortest paths in $c$ are more likely to include the same hyperedges.
As a consequence, the denser is $c$, the lower the number of hyperedges needed to cover all the shortest $s$-paths in $c$.
As shown in \Cref{sec:la}, the number of shortest paths covered by a hyperedge is a good indicator of its suitability as landmark.

\mpara{Sampling-based strategy.}
The sampling strategy assigns a probability $\mathbb{P}_{c^s} \in [0, 1]$ to each $s$-connected component $c^s$, and iteratively samples a connected component from $\textsf{CC}$, until the estimated oracle size $Q_{\textsf{est}}$ reaches the target size $Q$.
When an $s$-connected component $c^s$ is sampled, an additional landmark is assigned to $c^s$ only if the number of landmarks already assigned to $c^s$ is lower than $|c^s|$ (there cannot be more landmarks than hyperedges in $c^s$).
If the landmark is assigned to $c^s$, $Q_{\textsf{est}}$ is updated by adding $|c^s|$, because the oracle will store the distances from the landmark to every hyperedge in $c^s$.
Let $\alpha \geq 0$ be the importance factor of the $s$-component size,
$\beta \geq 0\, \text{ s.t. } \alpha + \beta \leq 1$ be the importance factor of the value $s$ for which the component is a $s$-connected component, and $1 - (\alpha + \beta)$ the importance factor of the number of vertices in the $s$-connected component.
Therefore, the larger is $\alpha$, the more important the size of the $s$-component over $s$ and its number of vertices.

Then, the sampling probability $\mathbb{P}_{c^s}$ of $c^s$ is
\begin{equation}
\mathbb{P}_{c^s} = \alpha\frac{|c^s|}{\zeta(\textsf{CC})} + \beta\frac{s}{\eta(\textsf{CC})} + (1 - \alpha - \beta)\frac{|V_{c^s}|}{\xi(\textsf{CC})}
\end{equation}
where \vspace{-\baselineskip}
\begin{align*}
% \label{eq:sampling}
\zeta(\textsf{CC}) = \sum\limits_{i=1}^{\smax}{\sum\limits_{j=1}^{|\textsf{CC}_i|}{|\textsf{CC}_{i,j}|}}\,,
\quad
\xi(\textsf{CC}) =& \sum\limits_{i=1}^{\smax}{\sum\limits_{j=1}^{|\textsf{CC}_i|}{|V_{\textsf{CC}_{i,j}}|}}\,,\\
\text{and} \quad \eta(\textsf{CC}) =& \sum_{i=1}^{\smax}{i*|\textsf{CC}_i|}.
\end{align*}

\begin{algorithm}[!t]
\footnotesize
\begin{algorithmic}[1]
\Require Connected Components $\textsf{CC}$, Max Oracle Size $Q$
\Require Min Dim $\dmin$
\Require Importance Factors $\Omega$
\Require Landmark Selection Strategy $\sigma$
\Ensure A set of landmarks $L[s]$ for each $s$
    \State $C^{\dmin} \gets $ connected components $c$ with $|c| > \dmin$
    \State $r_1 \gets C^{\dmin} \text{ ranked by decreasing size }$
    \State $r_2 \gets C^{\dmin} \text{ ranked by decreasing num of vertices }$
    \State $r_3 \gets C^{\dmin} \text{ ranked by decreasing s }$
    \State $r_4 \gets C^{\dmin} \text{ ranked by increasing landmarks assigned }$
    \State $Q_{\textsf{est}} \gets 0$; $L \gets \varnothing$
    \While{$Q_{\textsf{est}} < Q$}
        \State $\textsc{rank} \gets \Call{aggRanking}{r_1, r_2, r_3, r_4, \Omega}$
        \State take a random $c^s$ in the first position of $\textsc{rank}$
        \State $L[s] \gets L[s] \cup \Call{selectLandmark}{c^s, L, s, \sigma}$ \label{line:selL}
        \State $Q_{\textsf{est}} \gets Q_{\textsf{est}} + |c^s|$
        \If{$|c^s|$ landmarks have been assigned to $c^s$}
            \State remove $c^s$ from the rankings
        \Else
            \text{ update} $r_4$
        \EndIf
    \EndWhile
    \State \Return $L$
\end{algorithmic}
\caption{\textsc{selectLandmarks}}\label{alg:landmark_selection}
\end{algorithm}

\mpara{Ranking-based strategy.}
The ranking strategy creates 4 rankings of the connected components in $\textsf{CC}$, finds a consensus of the rankings, and then assigns landmarks to connected components according to the consensus.
\Cref{alg:landmark_selection} illustrates the procedure.
First, the algorithm creates a ranking by decreasing size ($r_1$), by decreasing number of vertices ($r_2$), by decreasing $s$ ($r_3$), and by increasing number of landmarks already assigned ($r_4$).
Then, it iteratively finds the consensus $\textsc{rank}$ via Procedure~\textsc{aggRanking} and selects a random component $c^s$ in the first position of $\textsc{rank}$, until the estimated oracle size $Q_{\textsf{est}}$ reaches the target size $Q$.
Similar to the sampling-based approach, $Q_{\textsf{est}}$ is updated by adding $|c^s|$. Then, $r_4$ is updated according to the new landmark assignment.
To avoid assigning more landmarks than the available hyperedges, when $|c^s|$ landmarks have been assigned to a component $c^s$, it is removed from all the rankings.
For efficiency, the selection of the landmarks via $\sigma$ is carried on iteratively, each time a new landmark is assigned to $c^s$ (\cref{line:selL}).

Procedure~\textsc{aggRanking} finds the consensus by solving an instance of \emph{Rank Aggregation with Ties}.
Rank Aggregation is the task of combining multiple rankings of elements into a single consensus ranking.
When the rankings are not required to be permutations of the elements, elements can placed in the same position (\emph{ties}).
The distance between two rankings with ties $\pi_1$ and $\pi_2$ is measured as the \emph{generalized Kendall-}$\tau$ \emph{distance} $K_{\tau,p}(\pi_1,\pi_2)$~\cite{kumar2010generalized},
which counts the pairs of elements for which the order is different in $\pi_1$ and $\pi_2$.

The \emph{optimal consensus} of a set of rankings $\Pi = \{\pi_1, \dots, \pi_k\}$ is the ranking $\pi$ that minimizes the weighted Kemeny score ($\textsc{WS}$)~\cite{kemeny1959mathematics},
i.e., the weighted sum of the generalized Kendall-$\tau$ distances between $\pi$ and each ranking in $\Pi$, where $\omega_i$ indicates the importance of $\pi_i$:
$\textsc{WS}(\pi, \Pi) = \sum\limits_{i=1}^k{\omega_i \cdot K_{\tau,p}(\pi_i,\pi)}$\,.

Several approaches have been proposed to find the optimal consensus. A recent survey~\cite{brancotte2015rank} shows that BioConsert~\cite{cohen2011using} is the best approach in many case scenarios.
Since solving optimal consensus is \textbf{NP}-hard for $k \geq 4$~\cite{betzler2008fixed}, BioConsert follows a local search approach that can result in sub-optimal solutions.
The algorithm iterates over each ranking $\pi_i \in \Pi$, applies operations to $\pi_i$ to make it closer to a median ranking, and finally returns the best one.

In our implementation,
we assign the same importance $\alpha \in [0, 1]$ to $r_1$ and $r_2$, and importance $\beta \in [0, 1]$ to $r_3$.
The importance of $r_4$ is set to $1$, to reduce the impact of the component sizes on the consensus, hence avoiding assigning landmarks always to the same large connected components.

\subsection{Landmark Selection}\label{sec:la}

\begin{figure*}[th]
    \centering
    \includegraphics[width=.9\linewidth]{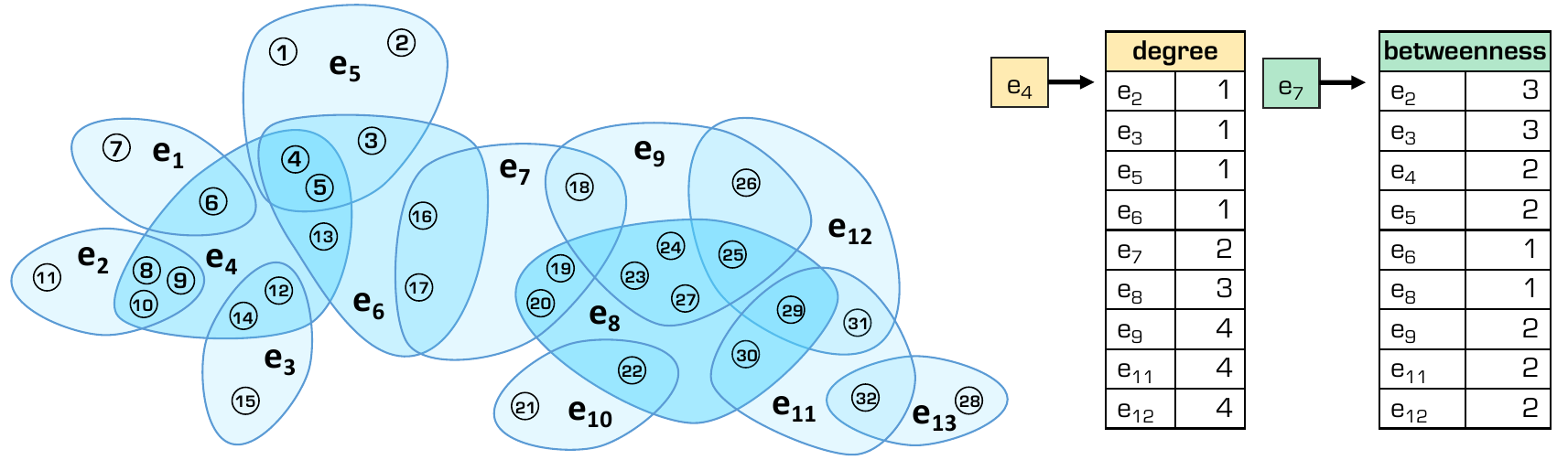}
    \caption{\rev{A hypergraph (left) and two $2$-distance oracles (right) with $1$ landmark selected using the \textsf{degree} ($e_4$) and the \textsc{betweenness} ($e_7$) landmark selection strategy. The oracles include the $2$-distances from the landmark to all the hyperedges reachable via $2$-paths.}}
    \label{fig:toy2}
\end{figure*}

Given a set of landmarks $L$, the $s$-distance $\hdist{s}(e,f)$ between two hyperedges $e$ and $f$ is upper-bounded by $\mathrm{ub}_s(e,f) = \min_{l \in L}(\hdist{s}(l,e) + \hdist{s}(l,f))$.
The upper bound is tight when there exists a landmark $l$ lying on a shortest $s$-path between $e$ and $f$ $\mathrm{ub}_s(e,f) = \hdist{s}(e,f)$.
As a consequence, better approximations can be obtained by selecting landmarks belonging to as many shortest $s$-paths as possible.
The optimal set of landmarks is the \emph{2-hop s-distance cover}.
A set of landmarks $L$ is a $2$-hop $s$-distance cover iff for any two hyperedges $e$ and $f$, $\hdist{s}(e,f) = \mathrm{ub}_s(e,f)$, and $L$ is minimal.
However, finding a 2-hop distance cover is an \textbf{NP}-hard problem~\cite{potamias2009fast}.
In the graph setting, more efficient landmark selection strategies have been proposed.
Many of these can be adapted to the hypergraph setting to select landmarks within an $s$-connected component $c^s$.

\spara{Random.} The \textsf{random} strategy is a baseline used to evaluate the effectiveness of the heuristics used by the more advanced strategies.
It selects the landmarks uniformly at random from $c^s$.

\spara{Degree.} Let $\Gamma_s(e)$ denote the set of \emph{s-neighbors} of $e \in c^s$, i.e., the hyperedges $f \in c^s$ such that $|e \cap f| \geq s$.
The cardinality of this set is called the \emph{s-degree} of $e$.
The \textsf{degree} \rev{(a.k.a. \textsf{deg})} strategy selects the hyperedges with largest $s$-degree.

\spara{Farthest.} The \textsf{farthest} \rev{(a.k.a. \textsf{far})} strategy was originally proposed by \cite{goldberg2005computing} for graphs.
The first landmark $l_1$ is selected uniformly at random from $c^s$.
Then, iteratively, the distances from the current set of landmarks to all the reachable hyperedges are computed.
The next landmark is selected from $c^s$ such that it is the most distant hyperedge from the landmarks previously selected.
Graph-based approaches usually assume graph connectivity~\cite{potamias2009fast}; however, as the value of $s$ increases, the number of $s$-connected components in the hypergraph also increases. Therefore, it is crucial to account for this disconnectivity when implementing this strategy for the hypergraph setting.

\spara{BestCover.} The \textsf{bestcover} \rev{(a.k.a. \textsf{bc})} strategy was proposed by \cite{tretyakov2011fast} for graphs, with the goal of approximately cover as many shortest paths as possible.
In the first phase, a set of hyperedge pairs is sampled from $c^s$, and the shortest $s$-paths between them are computed.
In the second phase, landmarks are iteratively chosen, such that they cover most of the shortest $s$-paths not yet covered by previously selected landmarks.
% If all the paths are covered but it needs more landmarks, it adopts the degree strategy.

\spara{Betweenness.} Similar to the \textsf{bestcover} strategy, the \textsf{between} \rev{(a.k.a. \textsf{bet})} strategy works in two phases.
In the first phase, it finds all the shortest $s$-paths between pairs of hyperedges sampled from $c^s$.
Then, it selects the hyperedges with highest betweenness centrality, i.e., the hyperedges that lie on most of the shortest $s$-paths found.
% If it needs more landmarks, it adopts the degree strategy.

\begin{example}
\rev{\Cref{fig:toy2} illustrates a hypergraph (left) and two $2$-distance oracles (right) with $1$ landmark selected using the \textsf{degree} ($e_4$) and the \textsf{between} ($e_7$) landmark selection strategies, respectively.
The oracles include the $2$-distances from the landmark to all the hyperedges reachable via $2$-paths.
In this particular case, there is a giant $2$-connected component consisting of all the hyperedges but $e_1$, $e_{10}$, and $e_{13}$, which form separate singleton components.
Therefore, each oracle stores the distances to $9$ out of $13$ hyperedges in the hypergraph.
Note that while the $1$-distance between $e_7$ and $e_9$ is $1$, their $2$-distance is $2$, because the overlap between $e_7$ and $e_9$ is lower than $2$, and thus they can reach each other only through $e_8$.
The \textsf{degree} oracle estimates the $2$-distance between $e_2$ and $e_6$ as
\begin{align*}
& \hest{2}(e_2, e_6) = {\mathrm{lb}_2(e_2, e_6) + \mathrm{ub}_2(e_2, e_6) \over 2} = \\
& {\left|\hdist{2}(e_4, e_2) - \hdist{2}(e_4, e_6)\right| + \hdist{2}(e_4, e_2) + \hdist{2}(e_4, e_6) \over 2} = 1.
\end{align*}
The \textsf{between} oracle estimates this distance as
\begin{align*}
{\left|\hdist{2}(e_7, e_2) - \hdist{2}(e_7, e_6)\right| + \hdist{2}(e_7, e_2) + \hdist{2}(e_7, e_6) \over 2} = 3.
\end{align*}
The actual distance is $\hdist{2}(e_2, e_6) = 2$.}
\end{example}

\subsection{Complexity Analysis}\label{sec:complexity}
\algo stores three kinds of information: the $s$-distance oracles, the sizes of the $s$-connected components, and the hyperedge-to-connected-component memberships.
The space complexity is thus upper-bounded by \rev{$O(Q + \smax |E| + \smax |E|) = O(Q + \smax |E|).$}

The running time of \algo is determined by the time required to \textbf{(i)} find the $s$-connected components, \textbf{(ii)} assign the landmarks to the $s$-connected components, and \textbf{(iii)} populate the $s$-distance oracles.
The first step requires up to \rev{$O(\smax |E| \cdot A(\smax |E|^2, |E|))$}, where $A$ is the Ackermann function.
The complexity of the second step depends on the landmark assignment and selection strategy used.
For the ranking-based strategy equipped with \textsf{degree}, the running time is bounded by \rev{$O(\smax (|E| + Q |E|^2 + L_Q |E|\log(|E|)))$}; and for the sampling-based strategy equipped with \textsf{degree} is bounded by \rev{$O(\smax (|E| + Q \log(|E|) + L_Q |E|\log(|E|)))$}. Here $L_Q$ is the number of landmarks selected.
The third step requires up to \rev{$O(\smax |E|^2)$}, because, in the worst case, the number of pairs of overlapping hyperedges is $|E|^2$.
Therefore the time complexity is \rev{$O(\smax |E| (A(\smax |E|^2, |E|) + Q |E|))$}.
% \enlargethispage{\baselineskip} 
% !TEX root = ../main.tex
\section{Experimental Evaluation}\label{sec:experiments}

The experimental evaluation of \algo has three goals.
First, we discuss our design choices and the impact of each parameter on the performance of \algo.
Then, we compare the performance of \algo with those of several baselines. %, in terms of four metrics.
Finally we showcase how \algo can be used to \textbf{(i)} devise a recommender system, and \textbf{(ii)} approximate the closeness centrality of proteins in PPI networks.

\spara{Datasets.}
Table~\ref{tbl:datasets} reports the characteristics of all the real-world hypergraphs considered in the experimental evaluation.
The data was obtained from \cite{benson2018simplicial,preti2021strud}.\footnote{ \url{http://www.sociopatterns.org/datasets}} \footnote{\url{http://konect.cc/networks}} \footnote{\url{https://snap.stanford.edu/biodata}}

\begin{table}[t]
    \centering
    \caption{Characteristics of the datasets: \rev{number of nodes, number of hyperedges, maximum size of a hyperedge, number of hyperedges of size greater than $2$, and number of hyperedges of size greater than $3$}.}
    \label{tbl:datasets}
    	\begin{tabular}{lrrrrr}
            \toprule
            \textbf{Dataset} & $\mathbf{|V|}$ & $\mathbf{|E|}$ & $\mathbf{d}$ & $\mathbf{|E_{2}|}$ & $\mathbf{|E_{3}|}$\\
            \midrule
            \textbf{NDC-C} & \num{1.1}K & \num{1}K & \num{24} & \num{1}K & \num{750}\\
			\textbf{ETFs} & \num{2.3}K & \num{2}K & \num{19} & \num{1.6}K & \num{985}\\
			\textbf{High} & \num{327} & \num{8}K & \num{5} & \num{7.8}K & \num{2.3}K\\
			\textbf{Zebra} & \num{10}K & \num{10}K & \num{62} & \num{4.9}K & \num{3.7}K\\
			\textbf{NDC-S} & \num{5.3}K & \num{10}K & \num{25} & \num{6.3}K & \num{5.1}K\\
			\textbf{Primary} & \num{242} & \num{13}K & \num{5} & \num{12704} & \num{4956}\\
			\textbf{Enron} & \num{50}K & \num{106}K & \num{64} & \num{104}K & \num{68}K\\
			\textbf{Epinions} & \num{756}K & \num{107}K & \num{162}K & \num{73}K & \num{60}K\\
			\textbf{IMDB} & \num{1}M & \num{267}K & \num{10} & \num{257}K & \num{245}K\\
          	\textbf{DBLP} & \num{1.9}M & \num{2.4}M & \num{17} & \num{2.2}M & \num{1.5}M\\
          	\textbf{Threads-SO} & \num{2.7}M & \num{7}M & \num{22} & \num{5.2}M & \num{1.3}M\\
          	\midrule
          	\textbf{chem-gene} & \num{2.3}K & \num{2.5}K & \num{147} & \num{1.7}K & \num{1.2}K\\
			\textbf{dis-function} & \num{16}K & \num{3.4}K & \num{6.3}K & \num{3.4}K & \num{3.4}K\\
			\textbf{dis-gene} & \num{17}K & \num{3.9}K & \num{6.2}K & \num{2.2}K & \num{2}K\\
			\textbf{dis-chemical} & \num{1.7}K & \num{4.7}K & \num{1.1}K & \num{4.4}K & \num{4.2}K\\
            \bottomrule
        \end{tabular}
\end{table}

\spara{Experimental environment.}
\algo is written in Java 1.8.
Code and datasets are available at \url{https://github.com/lady-bluecopper/HypED}.
We run the experiments on a $24$-Core ($1.90$ GHz) Intel(R) Xeon(R) E5-2420 with 126GB of RAM, Ubuntu $18.04$, using all the cores available.
We set $\smax = 10$ and $\dmin = 4$ for all experiments.
For convenience, we set $Q = \ell \cdot |E|$, and pass the parameter $\ell$ to
\Cref{alg:create_oracle} instead of $Q$.
As \algo involves a probabilistic component, we execute each experiment $10$ times with different seeds and report averages.
Finally, the query hyperedges are randomly selected via a stratified sampling algorithm, which ensures that the set of queries includes hyperedges connected via $s$-paths for $s \in [1, 10]$.
Recall that the higher $s$, the lower the probability that two hyperedges are in the same $s$-connected component.
Hence, a naive sampler might select too few hyperedges connected via higher-order paths.

\spara{Metrics.}
We evaluate the performance of the algorithms according to four metrics.
The lower their value, the better the algorithm.
\textbf{OFF} indicates the time required to create the data structures (e.g., indices or oracles) used at query time;
\textbf{TimeXQ} is the average time required to answer a distance query;
\textbf{MAE} is the mean absolute error of the distance estimates;
\textbf{RMSE} is the root mean squared error of the distance estimates.
Furthermore, in some analyses, we show the distribution of the $L_1$-norm of the differences between real distances and estimates.

We note that accuracy bounds of distance oracles have been proven only for connected graphs~\cite{baswana2009all}.
For disconnected graphs, they are hard to obtain, unless some connectivity assumption is made. For this reason, we evaluate the accuracy of \algo only empirically.

\begin{figure}[t!]
	\centering
	\includegraphics[width=\columnwidth]{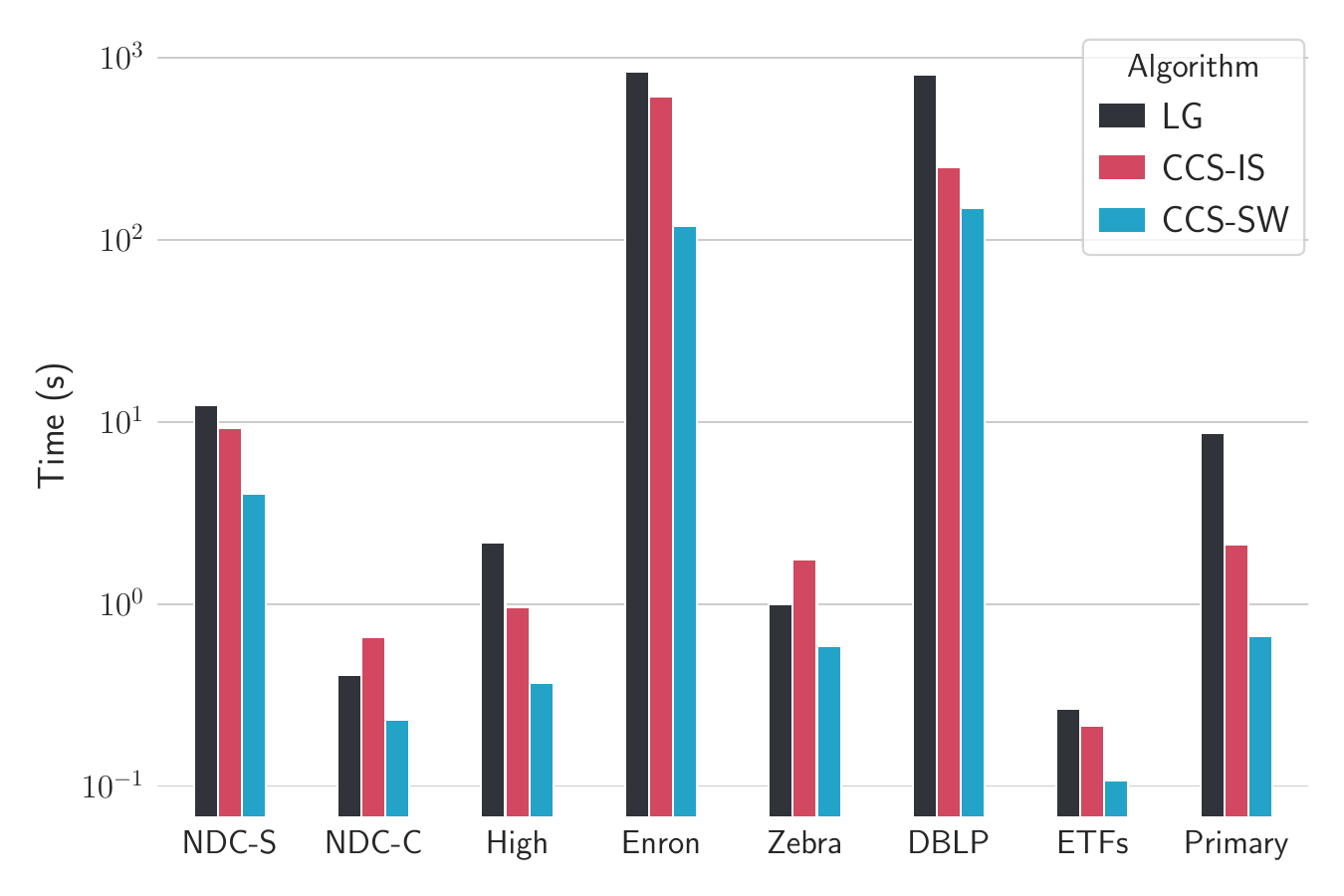}
	\caption{Time required to find the $s$-connected components for $s \in [1,10]$ using \lineg, \ccsbase, and \ccswise.}
	\label{fig:ccs_time}
\end{figure}

\spara{Baselines.}
We compare two versions of \Cref{alg:create_oracle}
(\algo and \algo-IS) with \textbf{(i)} an approximate baseline, \base, designed for hypergraphs, and \textbf{(ii)} two state-of-the-art approaches, \ctl~\cite{li2020scaling} and \hl~\cite{farhan2018highly}, designed for graphs.
We do not consider the method presented in \cite{qi2013toward}, as it is a distributed algorithm, while we focus on centralized algorithms.
The hypergraph-oriented algorithms are implemented in Java 1.8, while the graph-oriented algorithms are written in C++11.

\begin{table*}[t!]
	\centering
	\small
	\caption{Comparison between \samp and \rank: MAE, RMSE, oracle building time (OFF), \rev{and (average) number of landmark assigned (NL)}.}\label{fig:LA_comparison}
	\sisetup{table-format=2.3, table-alignment-mode=format}
	\begin{tabular}{l ccS[table-number-alignment=left]r ccS[table-number-alignment=left] S[table-number-alignment=right]}
		\toprule
		 & \multicolumn{4}{c}{\rank} & \multicolumn{4}{c}{\samp}\\
		 \cmidrule(lr){2-5} \cmidrule(lr){6-9}
		\textbf{Dataset} & \text{MAE} & \text{RMSE} & \text{OFF (s)} & \multicolumn{1}{c}{\rev{NL}} & \text{MAE} & \text{RMSE} & \text{OFF (s)} & \rev{AVG NL}\\
		\midrule
		ETFs	& 0.5980 & 1.5490 & 16.400 & \rev{1281} & 0.7489 & 1.6201 & 2.170 & \rev{356.9}\\
		High	& 0.9411 & 1.5703 & 14.537 & \rev{545} & 1.2167 & 1.8691 & 11.971 & \rev{103.6}\\
		NDC-C	& 0.4442 & 0.8564 & 2.856  & \rev{309} & 0.5150 & 0.8972 & 1.444 & \rev{147.5}\\
		NDC-S	& 0.7019 & 1.1286 & 103.830 & \rev{1559} & 0.8732 & 1.2312 & 38.706 & \rev{195.0}\\
		Primary	& 0.7881 & 1.1833 & 48.861 & \rev{649} & 0.9973 & 1.3118 & 39.560 & \rev{108.6}\\
		Zebra	& 1.3288 & 2.4162 & 95.767 & \rev{2328} & 1.6445 & 2.3963 & 4.850 & \rev{377.5}\\
		\bottomrule
	\end{tabular}
    \vspace{-1mm}
\end{table*}

\algo, \algo-IS, and \base create an $s$-distance oracle for each $s$ up to $\smax$, but they differ in how they populate them.
\base assigns landmarks with no regard to the connectivity of the hypergraph at the different values of $s$.
Given a target oracle size $Q$ and an estimated oracle size $Q_{\textsf{est}}$ initialized to $0$, it iterates until $Q_{\textsf{est}} \geq Q$.
At each iteration, it assigns $(Q - Q_{\textsf{est}})/(|E|*\smax)$ landmarks to each $s$-distance oracle, according to one of the landmark selection strategies.
\algo-IS assigns the landmarks to each $s$-distance oracle independently, while exploiting the information on the $s$-connected components.
Given a target oracle size $Q$, it ensures that each $s$-distance oracle has size $Q / \smax$, by calling Procedure~\textsc{selectLandmarks} with inputs $CCS = $ $s$-connected components, and $Q / \smax$ in place of $Q$.
This algorithm differs from \algo in that it treats all the oracles equally, and finds the $s$-connected components by using \ccsbase.
Since all the connected components given to the probabilistic strategy are $s$-connected components for the same value of $s$, $\beta$ is set to $0$.
Finally, \algo is simply \Cref{alg:create_oracle}.

\ctl~\cite{li2020scaling} improves the state-of-the-art 2-hop pruned landmark labeling approach (PLL~\cite{akiba2013fast}), by first decomposing the input graph in a large \emph{core} and a \emph{forest} of smaller trees, and then constructing two different indices on the core-tree structure previously generated.
A parameter $d$ bounds the size of the trees in the forest and regulates the trade-off between index size and query time.
We denote the algorithm with \ctl-$d$.

\hl~\cite{farhan2018highly} is a landmark-based algorithm that first selects a set of $l$ vertices, and then populates a \emph{highway} and a \emph{distance} index.
At query time, the algorithm first finds an upper-bound to the distance exploiting the highway index, and then, finds the distance in a sparsified version of the original graph.
We denote the algorithm with \hl-$l$.

\subsection{Performance Analysis of \algo}
We first compare the algorithm for computing the $s$-connected components against two baselines.
Then, we discuss the differences between the two procedures to assign landmarks.

\spara{Connected Components.}
We compare \Cref{alg:ccs} with two non-trivial baselines.
All the three strategies are based on the union-find algorithm, but they differ in how they exploit it.
The first baseline (\lineg) constructs the line graph $\mathcal{L}(H)$ by iterating over the hyperedges, exploiting a dynamic inverted index to efficiently find the number of vertices that each hyperedge shares with the other hyperedges. Then, it runs union-find for each $s$-line graph $\mathcal{L}_s$.
The second baseline, \ccsbase (\emph{Independent Stages}), finds the $s$-connected components for each $s$ separately.
Each time, it builds a dynamic inverted index to find the hyperedge $s$-overlaps, and then runs union-find using those overlaps.

\Cref{fig:ccs_time} shows the running time of the algorithms on all the datasets, for $s \in [1,10]$.
In most cases, the line-graph-based is the most expensive, as it must compute all the pairwise hyperedge overlaps.
Our technique,
that in \Cref{fig:ccs_time} is dubbed \ccswise, (\emph{StageWise}), outperforms the two baselines in all the datasets, thanks to the exploitation of \emph{(i)} the $(s+1)$-connected components for the initialization of the $s$-connected components, and \emph{(ii)} the transitive property for the reduction of the number of overlaps to compute.

\spara{Landmark assignment strategies.}
In the following experiments, we set $\ell = 30$ and select the landmarks using the \rev{\textsf{degree}, the \textsf{farthest}, and the \textsf{bestcover}} selection strategy.
For the sampling-based strategy (\samp), we consider the values $\{0.2, 0.33, 0.4\}$ for $\alpha$ and $\{0.6, 0.33, 0.2\}$ for $\beta$.
We pick such values with the goal of comparing the performance of \algo when \textbf{(i)} there is a higher probability of assigning a landmark to a higher-order connected component ($\alpha < \beta$), \textbf{(ii)} there is a higher probability of assigning a landmark to a larger connected component ($\alpha > \beta$), and \textbf{(iii)} the contribution of the component size and of $s$ is the same ($\alpha = \beta$).
We note that any other combination of values can be used as well, as long as $2 \cdot \alpha + \beta = 1$.

\begin{figure}[t!]
	\centering
	\includegraphics[width=\columnwidth]{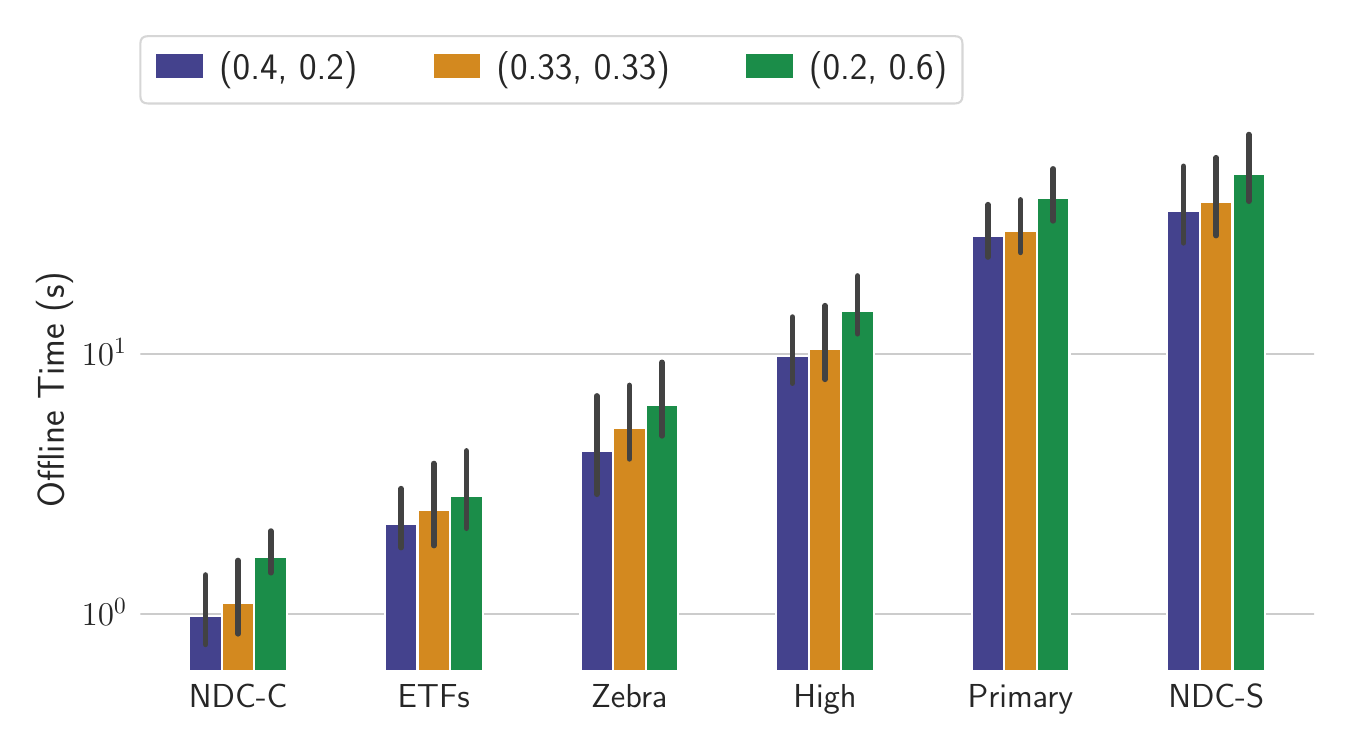}
	\includegraphics[width=\columnwidth]{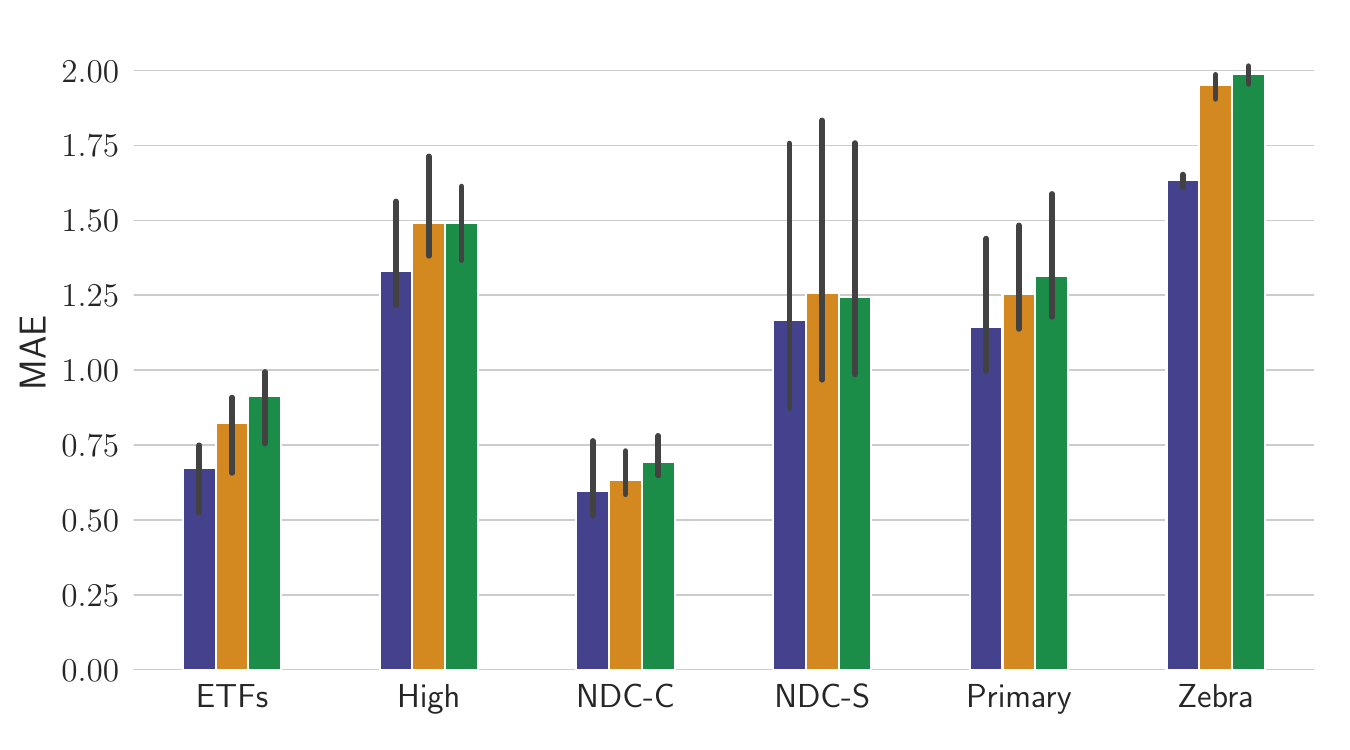}
	\caption{\rev{Oracle building time (top) and MAE (bottom) using the \samp landmark assignment strategy for different combinations of importance factors $(\alpha, \beta)$. The charts report mean and standard deviation among the values for the \textsf{degree}, \textsf{farthest}, and \textsf{bestcover} landmark selection strategies.}}
	\label{fig:time_mae_importance}
    \vspace{-3mm}
\end{figure}

\Cref{fig:time_mae_importance} shows that we can achieve more accurate estimates by giving more importance to the value of the parameter $s$, rather than the component size.
Indeed, higher values of $\alpha$ increase the probability of selecting landmarks within the same connected components (the largest ones), whereas larger values of $\beta$ increase the probability of selecting connected components of various sizes.
In fact, the latter configuration ensures the selection of landmarks in $s$-connected components for different values of $s$.
In turn, selecting landmarks in different $s$-connected components allows the oracle to answer a wider number $s$-distance queries, because at least one landmark is needed to obtain lower and upper bounds to the actual distance.
The answers, however, may be less accurate.
By looking at the mean MAE values, we observe lower errors when using the parameter combination $(\alpha, \beta) = (0.2, 0.6)$.

\begin{figure}[t!]
	\centering
	\includegraphics[width=\columnwidth]{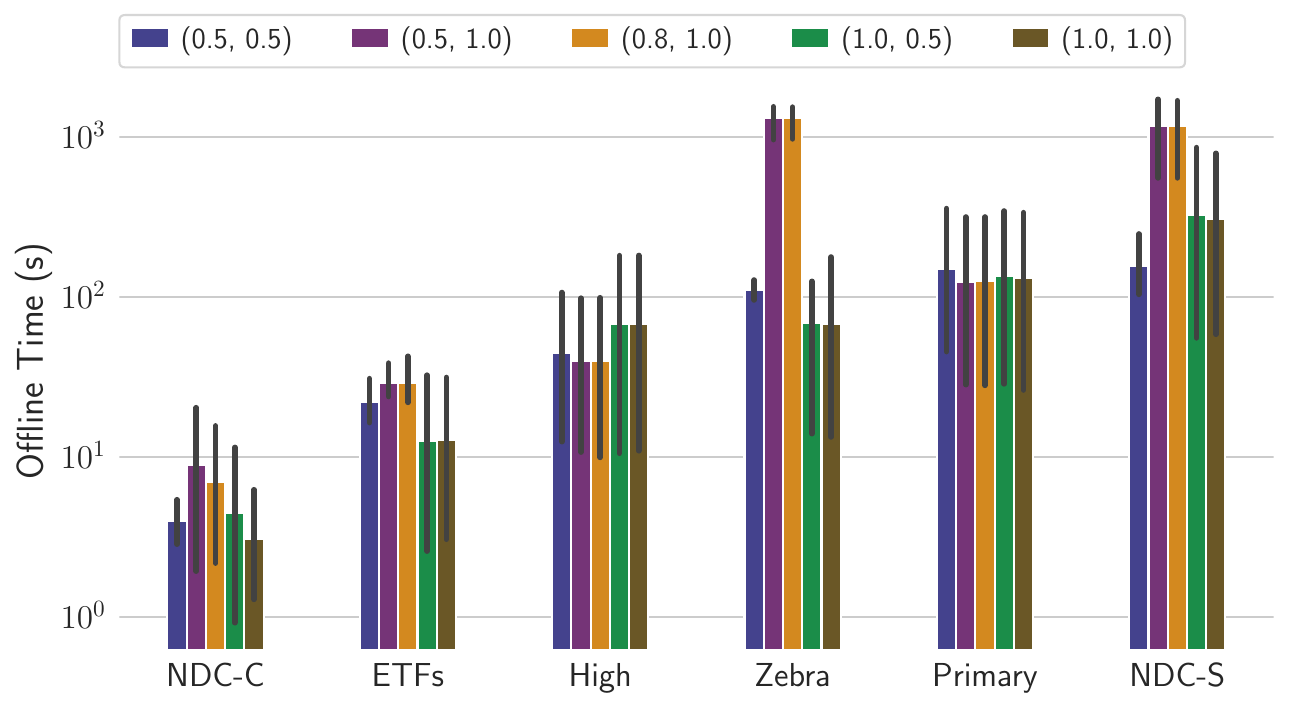}
	\includegraphics[width=\columnwidth]{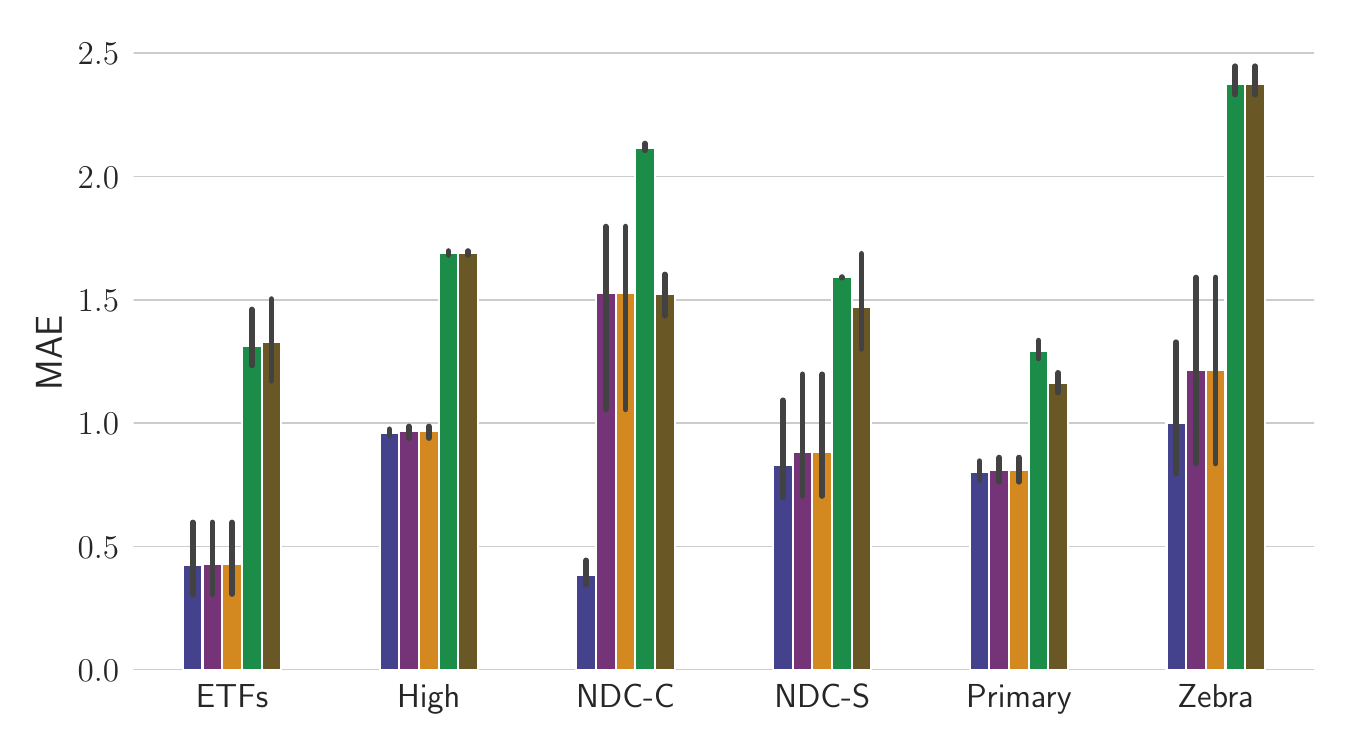}
	\caption{\rev{Oracle building time (top) and MAE (bottom) using the \rank landmark assignment strategy for different combinations of importance factors $(\alpha, \beta)$. The charts report mean and standard deviation among the values for the \textsf{degree}, \textsf{farthest}, and \textsf{bestcover} landmark selection strategies.}}
	\label{fig:time_mae_importance_rank}
    \vspace{-3mm}
\end{figure}

For the ranking-based strategy (\rank), we considered the values $\{0.5,$ $0.8,$ $1.0\}$ for both $\alpha$ and $\beta$.
When $\alpha$ is $1$, we observe worse MAE values \rev{(\Cref{fig:time_mae_importance_rank})}, because the probability of selecting landmarks in different connected components is higher when the importance factor of $r_4$ is strictly higher than the other factors.
In all other cases, differently from the sampling-based strategy, the performance are comparable, which means that the \rev{MAE values for the combinations $\alpha \leq \beta$ and $\alpha \geq \beta$ are nearly identical}.
\rev{By looking at the oracle building times, we select the combination $(\alpha, \beta) = (0.5, 0.5)$ as best performing combination.}

\Cref{fig:LA_comparison} reports a comparison between the sampling-based strategy with $(\alpha, \beta) = (0.2, 0.6)$  and the ranking-based strategy with $(\alpha, \beta) = (0.5, 0.5)$, which are the best performing variants of the two strategies.
It shows MAE, RMSE, OFF, \rev{and number of landmarks assigned}, for several datasets.

\begin{figure}[t!]
	\centering
	\includegraphics[width=\columnwidth]{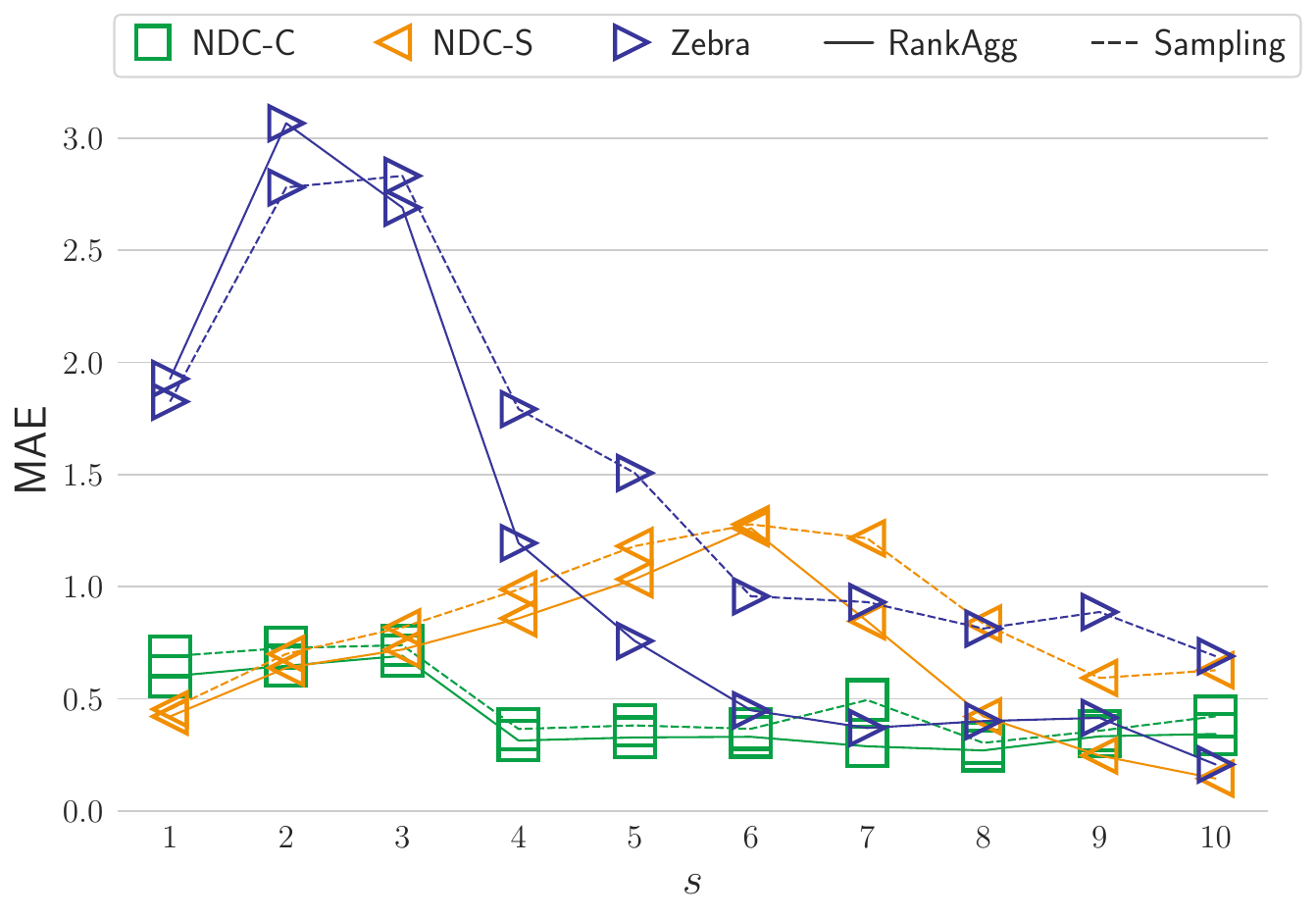}
	\caption{\rev{MAE of $s$-distance queries for different values of $s$, using the \samp and the \rank landmark assignment strategy and the \textsf{degree} landmark selection.}}
	\label{fig:mae_s_comparison}
\end{figure}

Due to the high time complexity of BioConsert, \rank shows worse running time than \samp.
Whereas the running time of \samp is affected by the hypergraph size and the number of connected components, that of \rank depends also on its past choices: each landmark assignment changes $r_4$, which, in turn, can complicate the task of finding a consensus.
This higher complexity comes with better results, though.
The MAE and RMSE \rev{consistently remain} lower than those of \samp, except for the RMSE in Zebra.
Indeed, \rank \rev{distributes} the landmarks more \rev{uniformly} across the connected components and, in particular, assigns them to smaller connected components. As a result, given a budget $Q$, it can assign more landmarks, and hence cover more connected components than \samp.
\rev{In the case of the largest connected components, typically found at smaller values of $s$, both strategies converge to a similar number of landmark assigned. This similarity results in similar approximate distance computations. Consequently, the MAE curves in \Cref{fig:mae_s_comparison} behave similarly for smaller $s$ values. However, a noticeable distinction emerges at larger $s$ values, because the corresponding $s$-connected components are smaller and thus better covered by \rank}.

\begin{figure}[t!]
	\centering
		\includegraphics[width=\columnwidth]{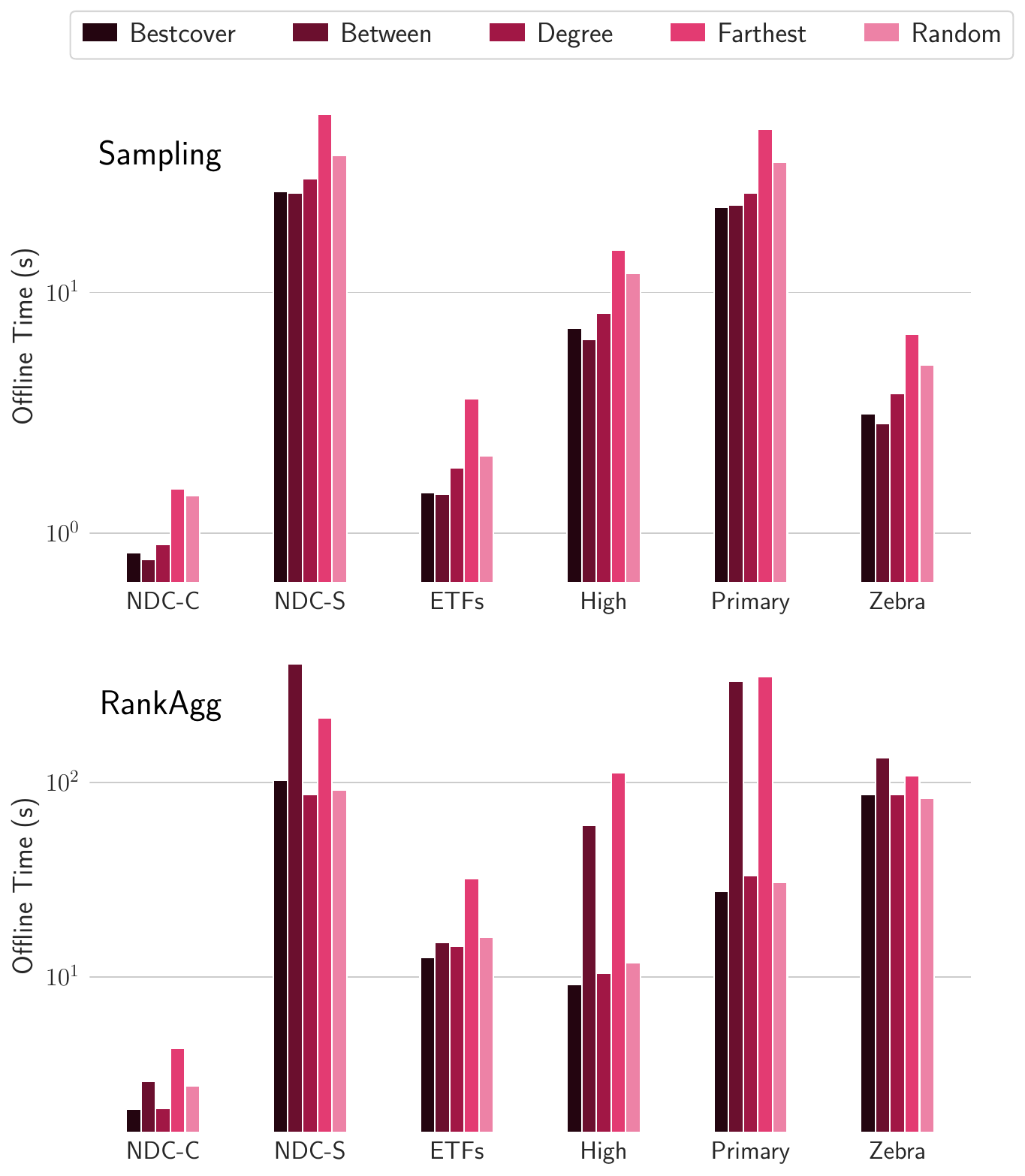}
	\caption{\rev{Time required to create the distance oracle when using the \samp (top) and the \rank landmark assignment strategy and different landmark selection strategies.}}
	\label{fig:LS_comparison_time}
\end{figure}

\begin{figure}[t!]
	\centering
		\includegraphics[width=\columnwidth]{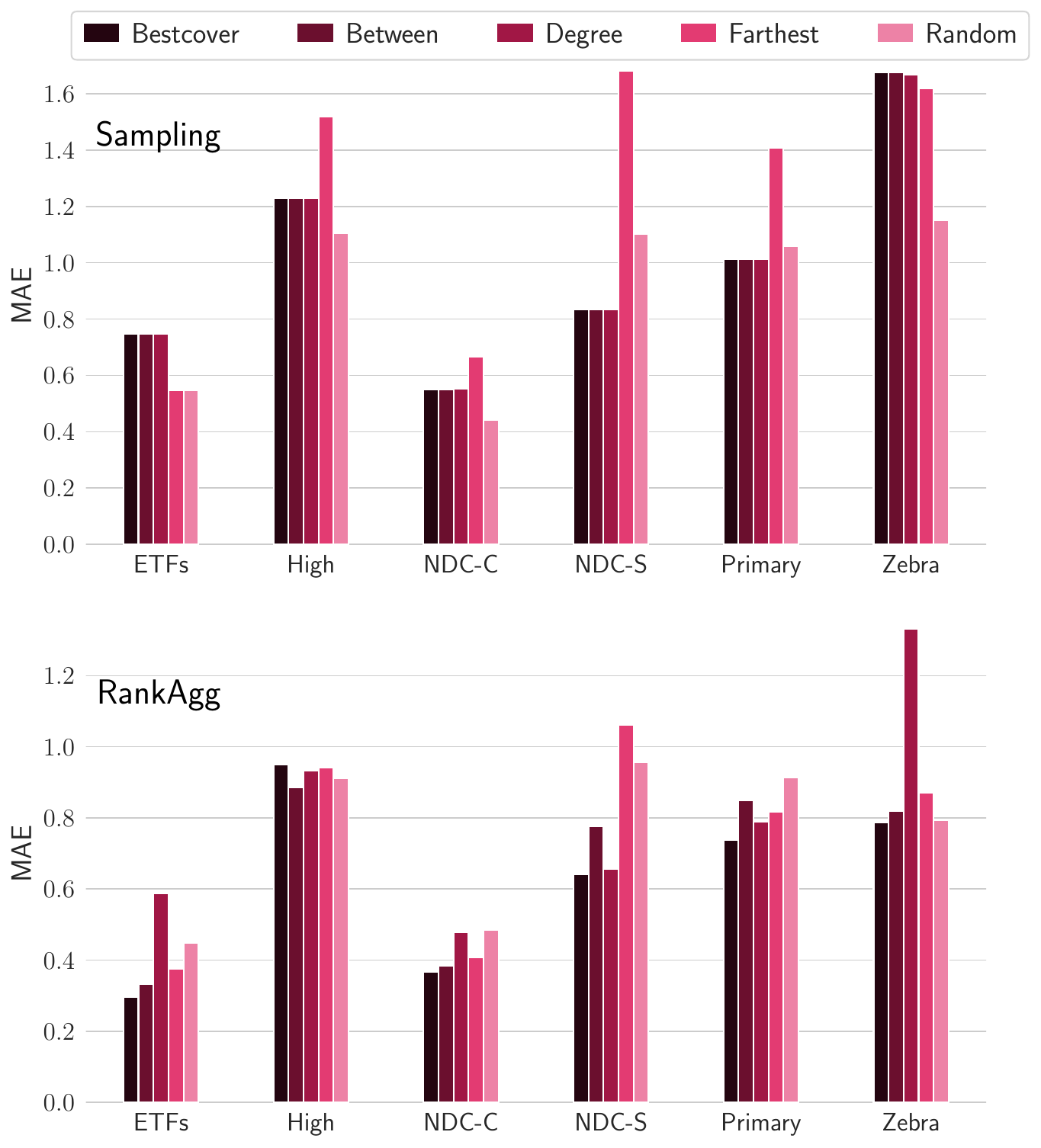}
	\caption{\rev{MAE values when using the \samp (top) and the \rank landmark assignment strategy and different landmark selection strategies.}}
	\label{fig:LS_comparison_mae}
\end{figure}

\begin{table}[t!]
	\centering
	\footnotesize
	\caption{MAE, RMSE, average query time, and average creation time, for \base, \algo-IS, and \algo.}
	\label{fig:comparison_ax}
\begin{tabular}{l cccS[table-number-alignment=left]}
	\toprule
	& \multicolumn{4}{c}{\base} \\
	\cmidrule(lr){2-5}
	\textbf{Dataset} & \text{MAE} & \text{RMSE} & \text{TimeXQ (\si{\us})} & \text{OFF (s)} \\
	\midrule
	ETFs & 1.6635 &	2.3319 & 0.882 & 2.4064 \\
	High & 1.5602 & 2.0992 & 0.947 & 10.1148 \\
	NDC-C & 1.1095 & 1.6988 & 0.482 & 1.1090 \\
	NDC-S & 1.3864 & 1.7209 & 0.436 & 54.2579 \\
	Primary & 1.3616 & 1.7490 & 0.887 & 35.2953 \\
	Zebra & 2.4091 & 2.7392 & 0.439 & 6.5455 \\
	\bottomrule
  \multicolumn{5}{c}{}\\
 & \multicolumn{4}{c}{\algo-IS}\\
	\cmidrule(lr){2-5}
	\textbf{Dataset} & \text{MAE} & \text{RMSE} & \text{TimeXQ (\si{\us})} & \text{OFF (s)} \\
	\midrule
	ETFs  & 1.0893 & 2.1026 & 0.387 & 0.4399 \\
	High & 1.1822 & 1.9806 & 0.691 & 3.0960 \\
	NDC-C & 0.5924 & 0.9937 & 0.422 & 1.0896 \\
	NDC-S & 0.8162 & 1.2194 & 0.425 & 33.8278 \\
	Primary & 0.8985 & 1.3510 & 0.677 & 10.2065 \\
	Zebra & 1.6687 & 2.5321 & 0.501 & 4.7281 \\
	\bottomrule
\multicolumn{5}{c}{}\\
& \multicolumn{4}{c}{\algo}\\
	\cmidrule(lr){2-5}
	\textbf{Dataset} & \text{MAE} & \text{RMSE} & \text{TimeXQ (\si{\us})} & \text{OFF (s)} \\
	\midrule
	ETFs  & 0.7437 &	1.6185 & 0.769 & 1.5461\\
	High  & 1.2167 & 1.8691 & 0.745 & 8.2114\\
	NDC-C & 0.5150 & 0.8972 & 0.408 & 0.8031\\
	NDC-S & 0.8732 & 1.2312 & 0.328 & 25.7270\\
	Primary &  0.9973 & 1.3118 & 0.685 & 26.0442\\
	Zebra & 1.6427 & 2.3959 & 0.374 & 4.2102\\
	\bottomrule
	\end{tabular}
    \vspace{-3mm}
\end{table}

\spara{Landmark selection strategies.}
We evaluate the performance of our algorithm when varying the landmark selection strategy.
We keep all the other parameters fixed: $\ell=30$, $\alpha=0.2$ and $\beta=0.6$ for \samp, \rev{and $\alpha = \beta = 0.5$ for \rank}.
For \textsf{bestcover} and \textsf{between}, we sample $40\%$ of the hyperedges to find the shortest paths.

\Cref{fig:LS_comparison_time} illustrates the OFF time \rev{for \samp (top) and for \rank (bottom), while \Cref{fig:LS_comparison_mae} displays the corresponding MAE values, for several datasets.}
\rev{For \samp}, the running time is comparable, with \textsf{random} and \textsf{farthest} being the slowest.
The running time of the former is affected by the poor selection of the landmarks (the population of the oracle for peripheral landmarks takes more time than that for central landmarks), while that of the latter is further increased by the procedure used to find all the hyperedges reachable from the landmarks already selected.
\rev{In the case of \rank, we observe a significantly higher running time also for \textsf{between}. This outcome can be attributed to \rank's tendency towards selecting a larger number of landmarks and to the higher complexity of \textsf{between} to select a landmark.}
By looking at the MAE, we can observe that, as expected, \textsf{bestcover} and \textsf{between} have similar performance, and are more frequently among the top-3 selection strategies.
Surprisingly, \textsf{random} reaches the top-3 in 4 out of 6 datasets \rev{for \samp}, proving the hardness of the landmark selection task in the case of very disconnected hypergraphs.

Since the difference in performance is relatively small and the running time of \samp is more predictable \textsc{and up to an order of magnitude lower than that of \textsc{RankAgg}}, the following experiments use \samp as landmark assignment strategy.

\begin{figure*}[t!]
\centering
	\begin{subfigure}{.45\linewidth}
		\centering
		\includegraphics[width=\columnwidth]{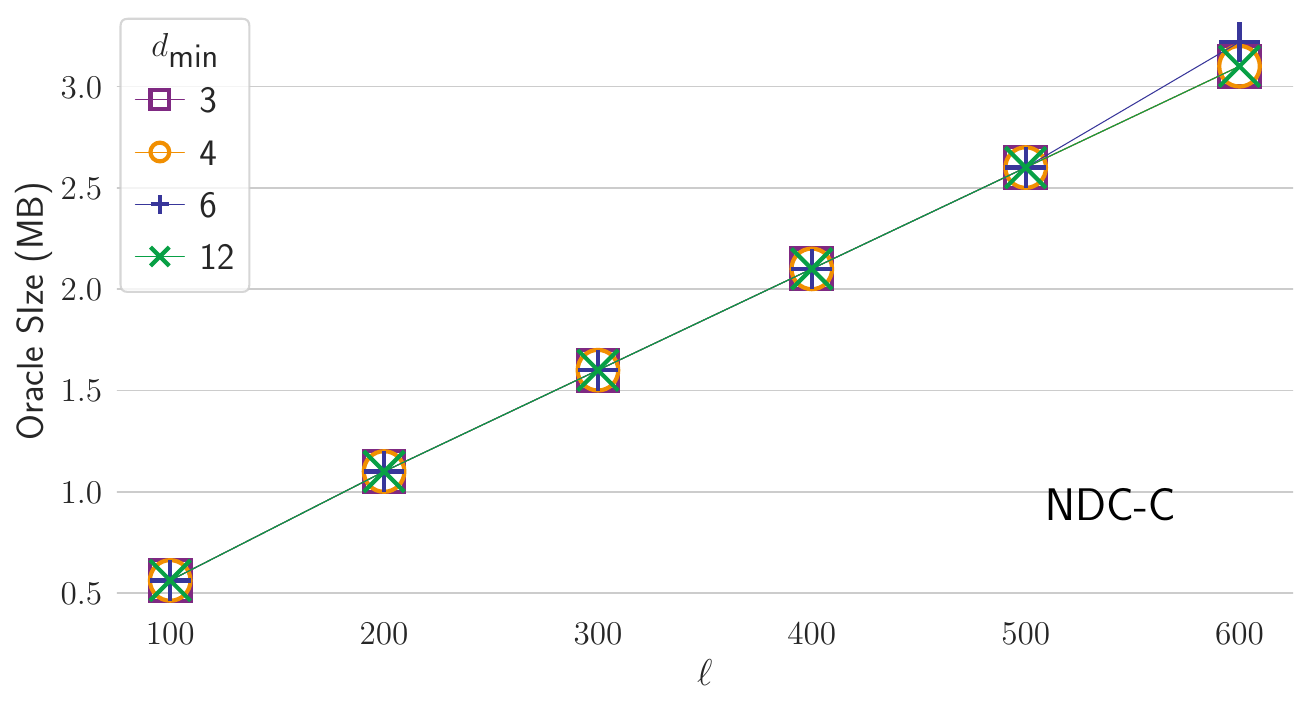}
	\end{subfigure}
	\begin{subfigure}{.45\linewidth}
		\centering
		\includegraphics[width=\columnwidth]{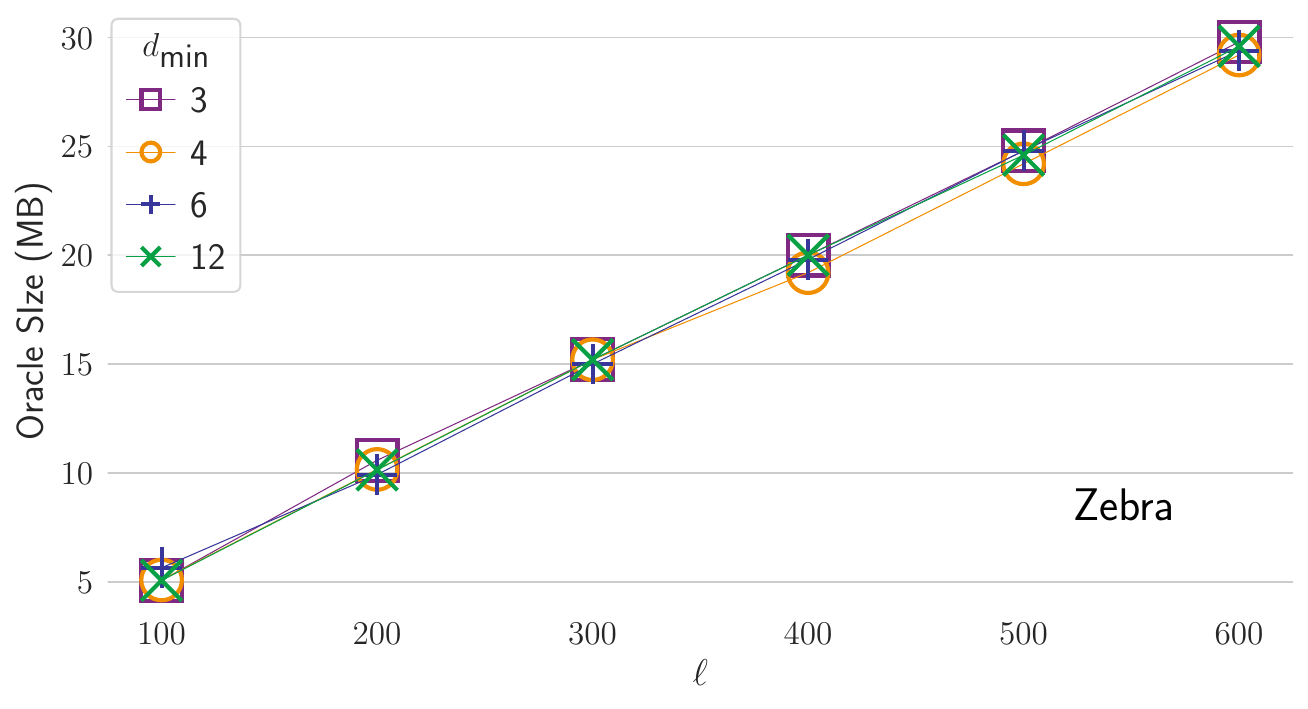}
	\end{subfigure}

	\begin{subfigure}{.45\linewidth}
		\centering
		\includegraphics[width=\columnwidth]{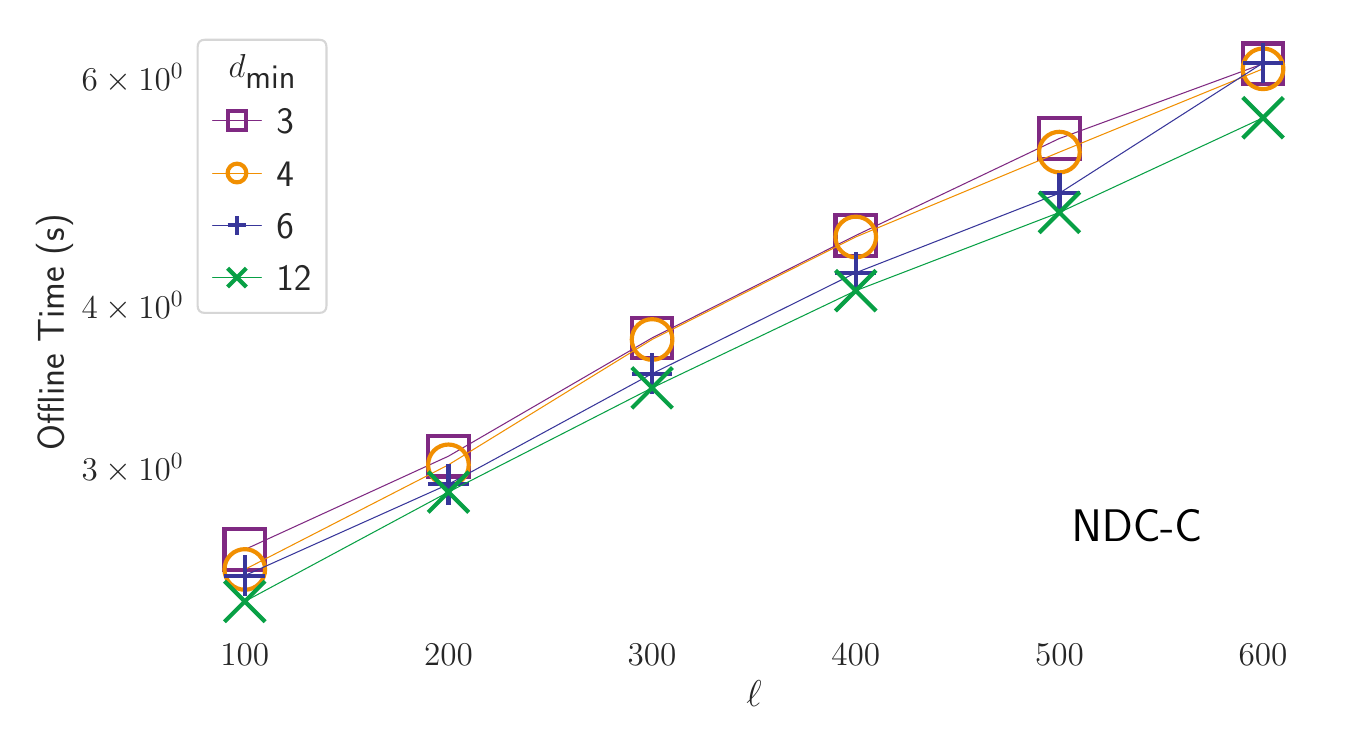}
	\end{subfigure}
	\begin{subfigure}{.45\linewidth}
		\centering
		\includegraphics[width=\columnwidth]{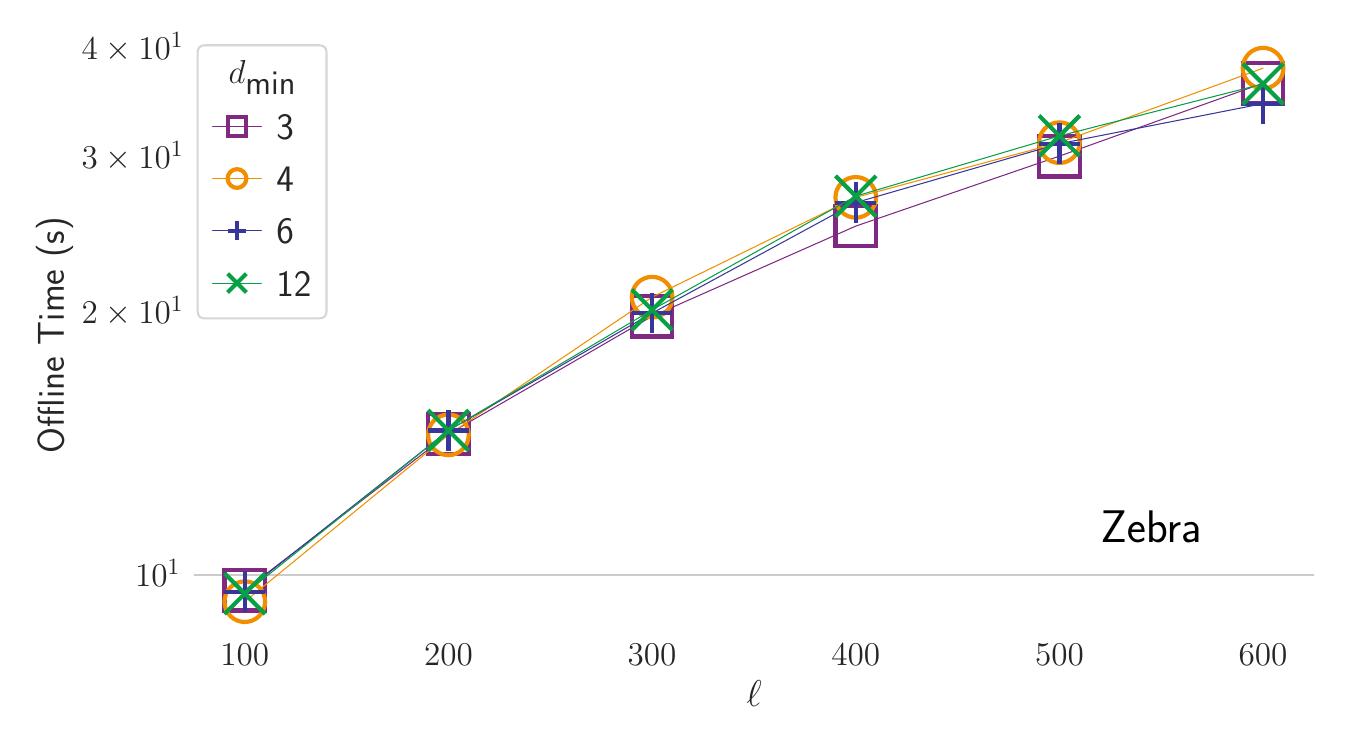}
	\end{subfigure}

	\begin{subfigure}{.45\linewidth}
		\centering
		\includegraphics[width=\columnwidth]{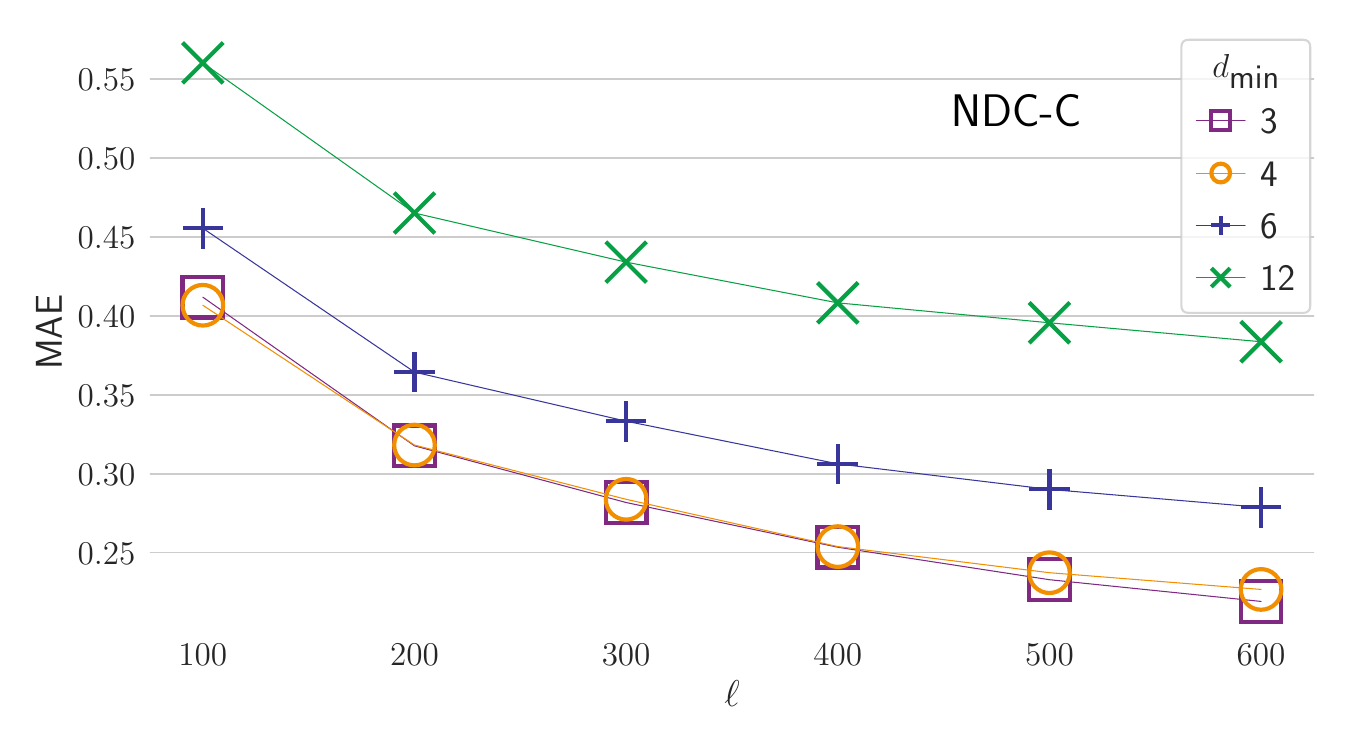}
	\end{subfigure}
	\begin{subfigure}{.45\linewidth}
		\centering
		\includegraphics[width=\columnwidth]{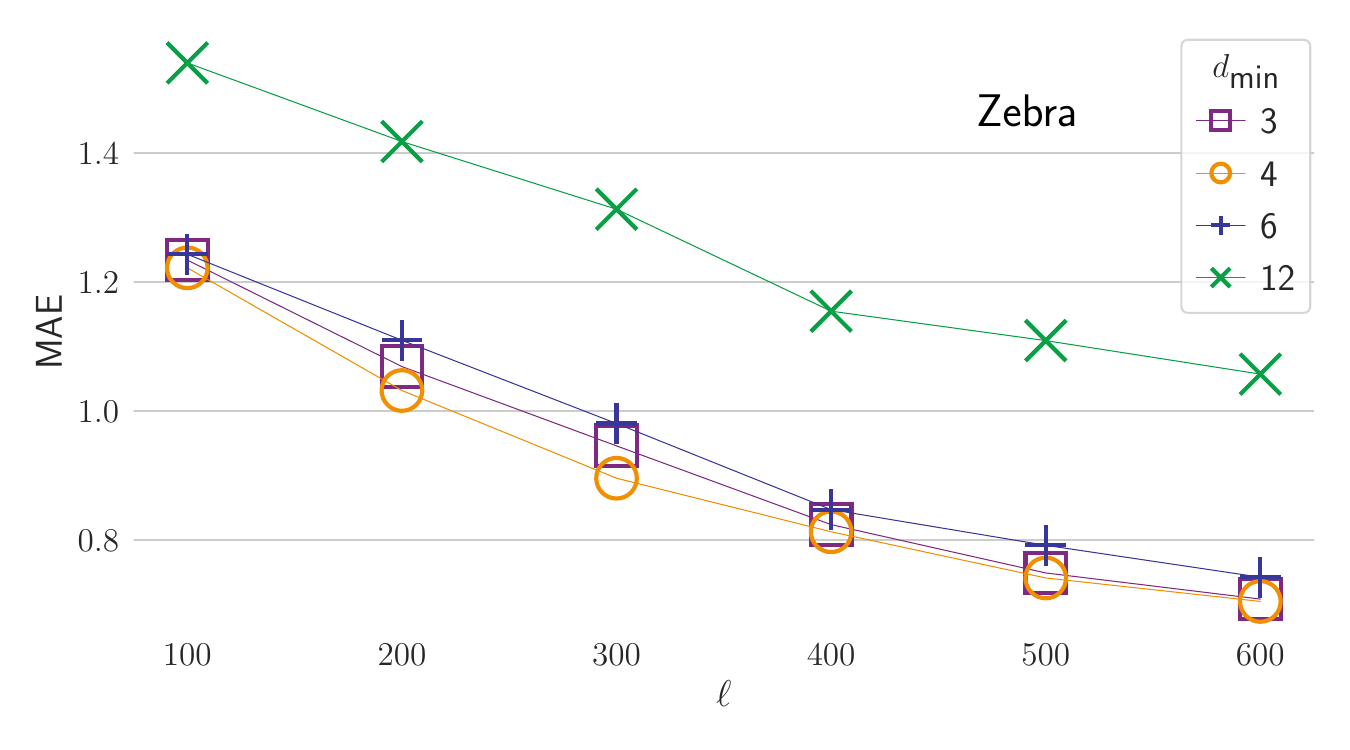}
	\end{subfigure}
	\caption{\rev{Oracle size (MB, top), oracle building time (s, middle), and MAE (bottom) for NDC-C (left) and Zebra (right), varying $\ell$ and $\dmin$, using the \samp landmark assignment strategy and the \textsf{degree} landmark selection strategy.}}
	\label{fig:q_d_zebra_ndcc}
\end{figure*}

\begin{figure*}[t!]
\centering
	\begin{subfigure}{.45\linewidth}
		\centering
		\includegraphics[width=\columnwidth]{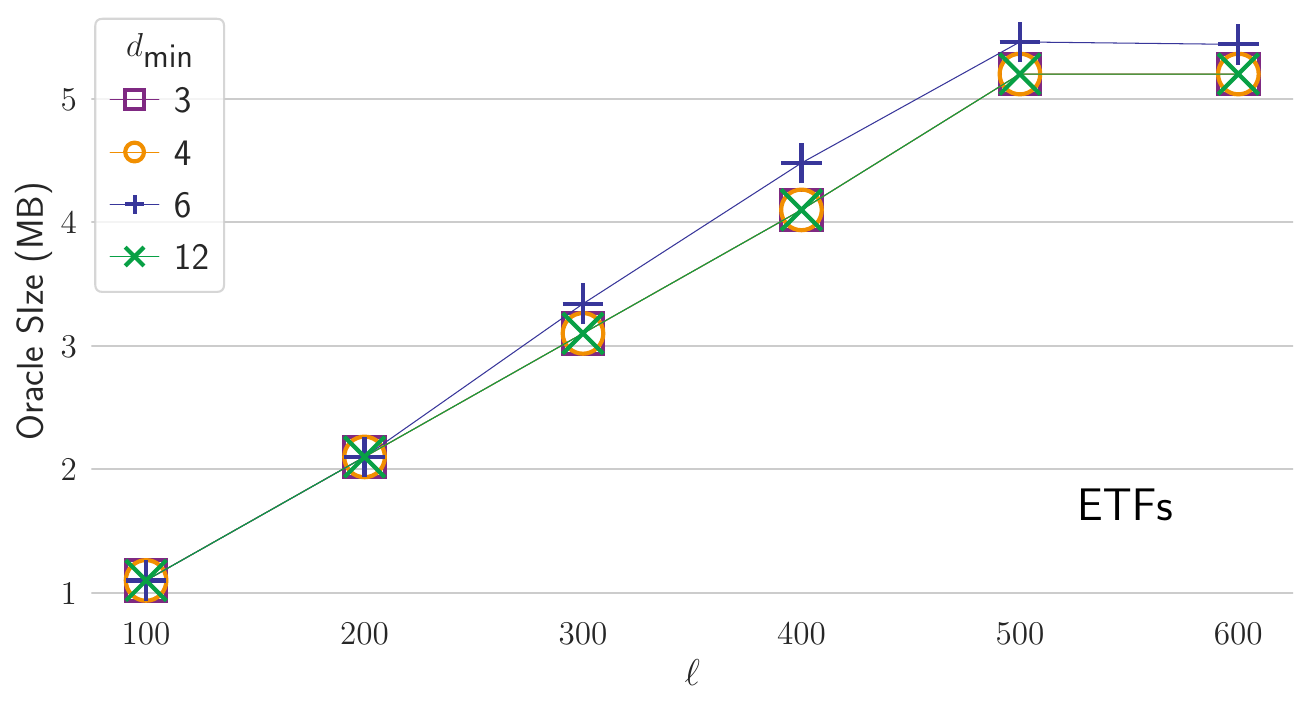}
	\end{subfigure}
	\begin{subfigure}{.45\linewidth}
		\centering
		\includegraphics[width=\columnwidth]{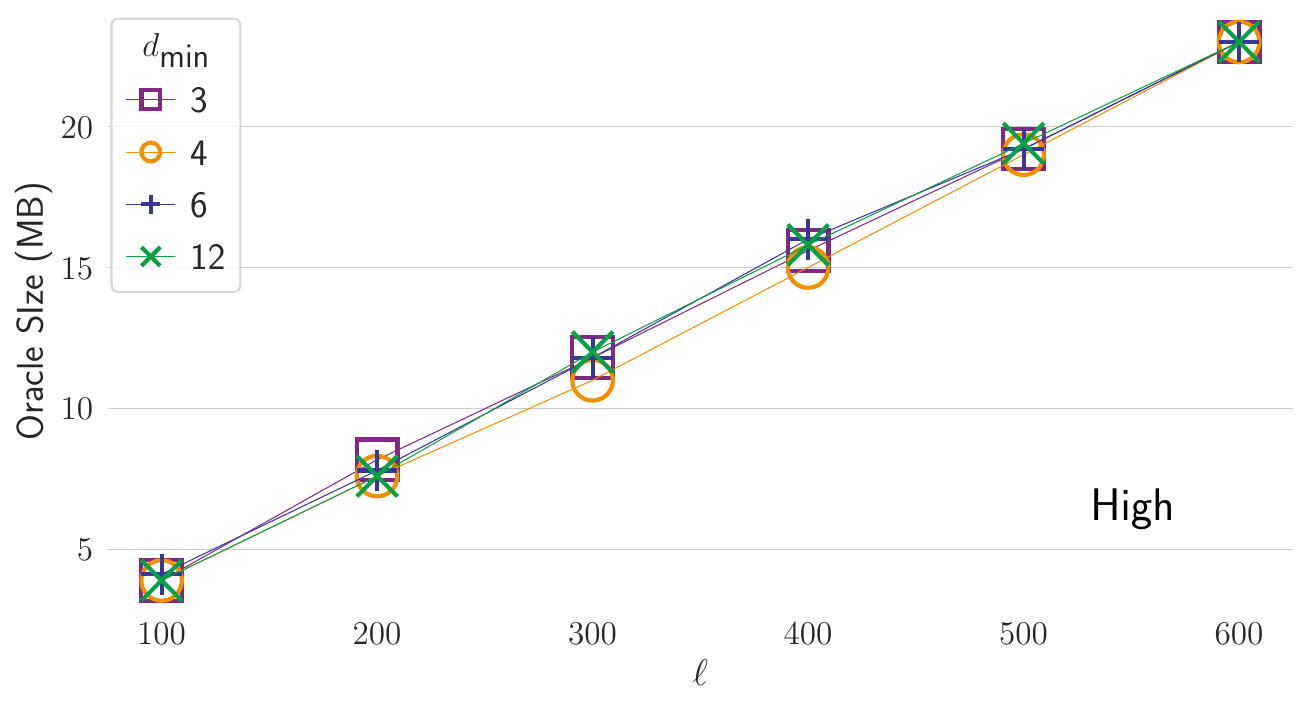}
	\end{subfigure}

	\begin{subfigure}{.45\linewidth}
		\centering
		\includegraphics[width=\columnwidth]{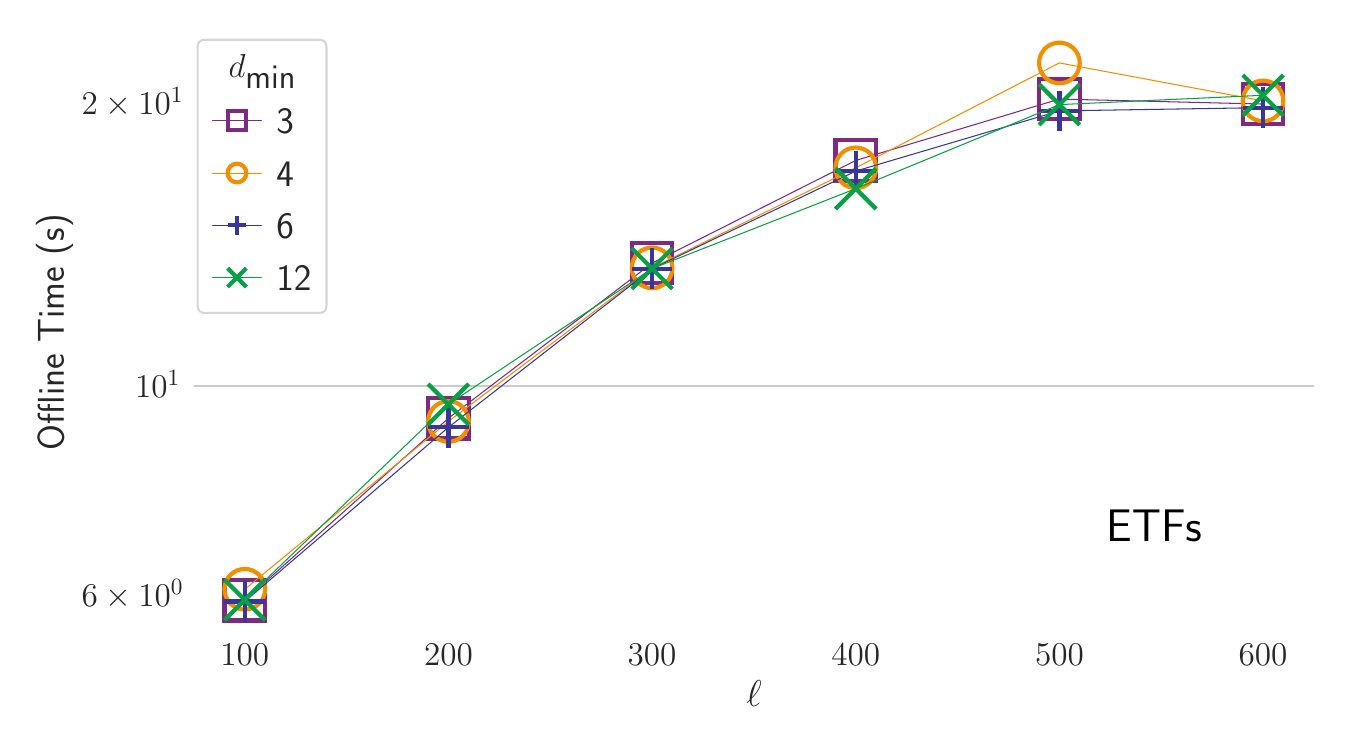}
	\end{subfigure}
	\begin{subfigure}{.45\linewidth}
		\centering
		\includegraphics[width=\columnwidth]{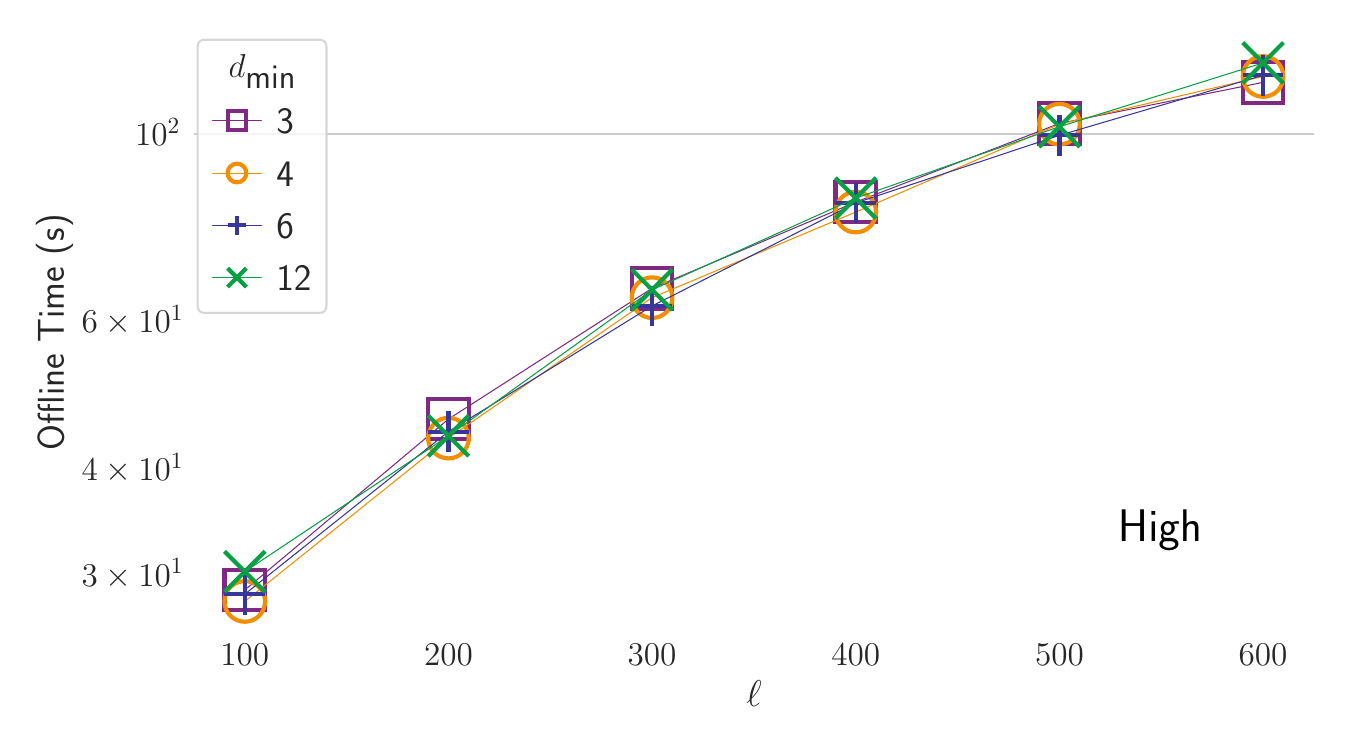}
	\end{subfigure}

	\begin{subfigure}{.45\linewidth}
		\centering
		\includegraphics[width=\columnwidth]{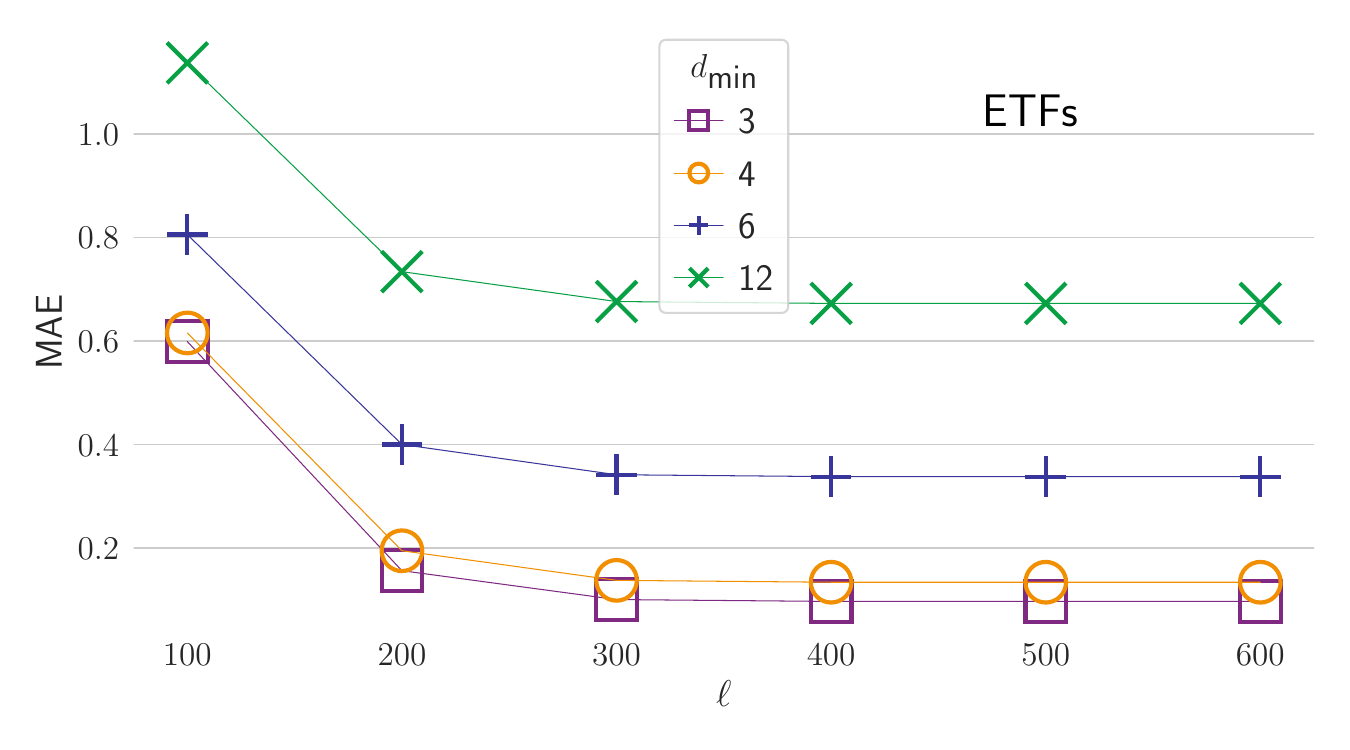}
	\end{subfigure}
	\begin{subfigure}{.45\linewidth}
		\centering
		\includegraphics[width=\columnwidth]{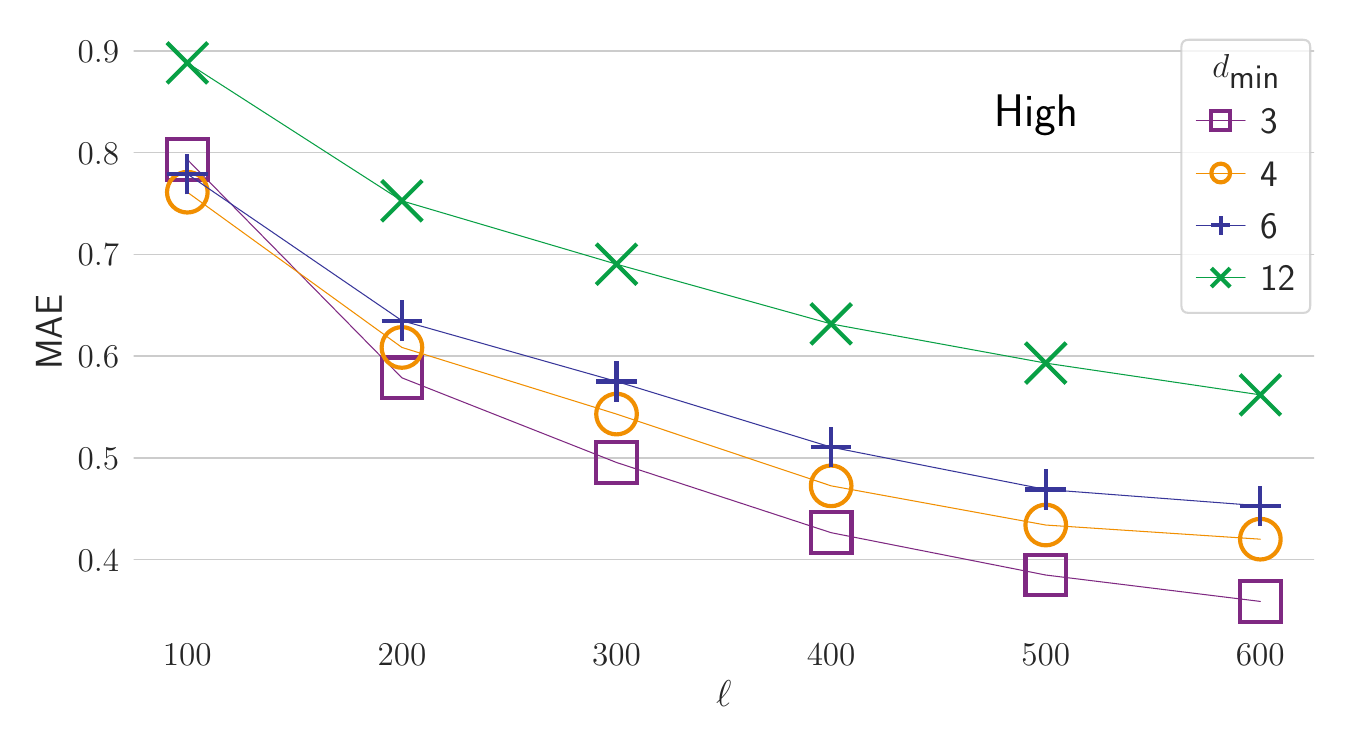}
	\end{subfigure}
	\caption{\rev{Oracle size (MB, top), oracle building time (s, middle), and MAE (bottom) for ETFs (left) and High (right), varying $\ell$ and $\dmin$, using the \samp landmark assignment strategy and the \textsf{degree} landmark selection strategy.}}
	\label{fig:q_d_etfs_high}
\end{figure*}

\rev{\spara{Desired oracle size and $\dmin$.}}
\rev{We investigate the influence of $\ell$ and $\dmin$ on both the size and accuracy of the oracle in $s$-distance approximations.
To this aim, we vary $\ell$ in $[100, \dots, 600]$ and $\dmin$ in $[3, 4, 6, 12]$.
We use the \samp landmark assignment strategy with $\alpha=0.2$ and $\beta=0.6$ and the \textsf{degree} landmark selection strategy.
\Cref{fig:q_d_zebra_ndcc} and \Cref{fig:q_d_etfs_high} illustrate the oracle sizes (in MB), the oracle building time (s), and the MAE for four datasets: NDC-C, Zebra, ETFs, and High.
We observe that $\dmin$ has minimal impact on the oracle size, as it essentially determines which $s$-connected components must be stored in the oracle and this information does not affect significantly the total size.
Similarly, it exerts little influence on the building time, serving mainly as a filter for connected components considered during the landmark assignment step.
However, there is a subtle effect on the MAE values. Allowing the assignment of landmarks to components of sizes $5$ and $6$ results in more accurate answers to a larger number of distance queries, consequently leading to lower MAE values.
}

\rev{As expected, the oracle size grows linearly with $\ell$. This growth, however, comes with an exponential surge in the running time.
In fact, each additional inclusion of a landmark requires the computation of all the distances from this landmark to every other hyperedge within the same $s$-connected component, a number that can escalate up to $|E_s|$ hyperedges.
In contrast, the improvement in MAE has a slow rise, with a significant inflection point around $\ell=200$. Across most datasets, we note low MAE values at $\ell=400$, suggesting a favorable trade-off between accuracy and computation time.}

\subsection{Comparison with Baseline}\label{subsec:comparison}

\begin{figure}[t!]
\centering
	\includegraphics[width=\columnwidth]{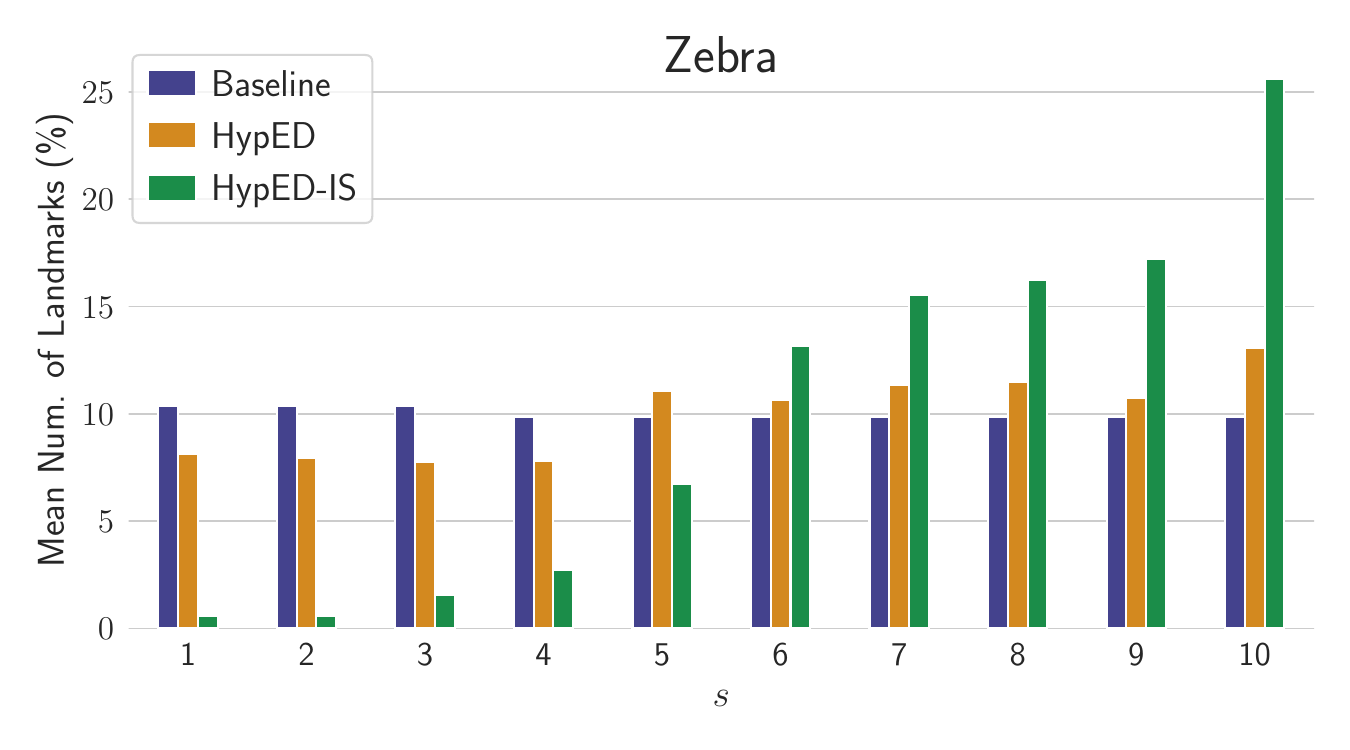}
	\caption{Mean number of landmarks selected by \base, \algo-IS, and \algo, for each $s$-distance oracle, in Zebra.}
	\label{fig:comparison_ax_lands}
\end{figure}

We first present the comparison among \algo, \algo-IS, and \base.
\Cref{fig:comparison_ax} and \Cref{fig:comparison_ax_lands} report the metrics for $\ell = 30$ and the {degree} landmark selection strategy.
The figure displays the results only for the dataset Zebra, but we observe similar behaviors in all the other datasets.
\base displays a higher OFF time, as it needs to perform several iterations before using up all the budget.
Conversely, \algo-IS and \algo have roughly the same creation time: \algo-IS takes more time to find the $s$-connected components, whereas the landmark assignment process of \algo is more complex.
While they tend to assign a similar number of landmarks, these algorithms differ in \emph{where} they assign them: \algo assigns roughly the same number of landmarks to each $s$-distance oracle, whereas \algo-IS assigns more landmarks to higher-order distance oracles.
This effect is a result of \algo-IS splitting the budget evenly among the $s$-distance oracles, and, at higher values of $s$, the hypergraph is more disconnected; thus, each landmark has a lower cost on the budget.
Similar to \algo, \base assigns the landmarks evenly among the $s$-distance oracles, but it fulfills the budget by using a lower number of landmarks, because it tends to assign landmarks to larger connected components.

Regarding the average query time, since \base is not aware of the connectivity of the hypergraph at various $s$, it takes more time to answer queries involving hyperedges either not connected or in the same small connected component. As a consequence its TimeXQ is higher.
In contrast, \algo-IS and \algo exploit the information on the $s$-connected components to assign the landmarks strategically, and hence they achieve both a lower TimeXQ and a lower MAE.
Even though \algo and \algo-IS display comparable MAEs in most of the datasets considered, we note that \algo is more versatile, because it always fully exploit the budget available.
In contrast, \algo-IS may construct oracles with size far lower than the desired value, due to the hypergraph being highly disconnected at larger values of $s$.
This is a consequence of the fact that \algo-IS divides the budget equally among the various $s$-distance oracles, before computing the $s$-connected components for each value of $s$.

\begin{figure*}[t!]
	\centering
	\begin{tabular}{c}
		\includegraphics[width=\textwidth]{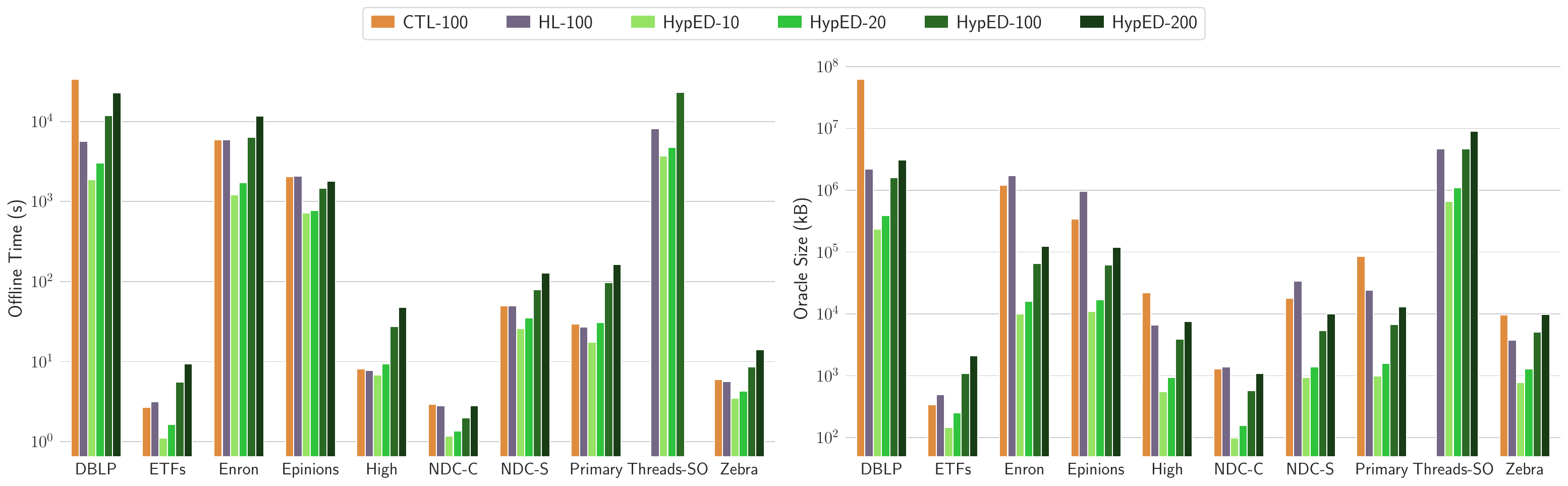}\\
		\includegraphics[width=\textwidth]{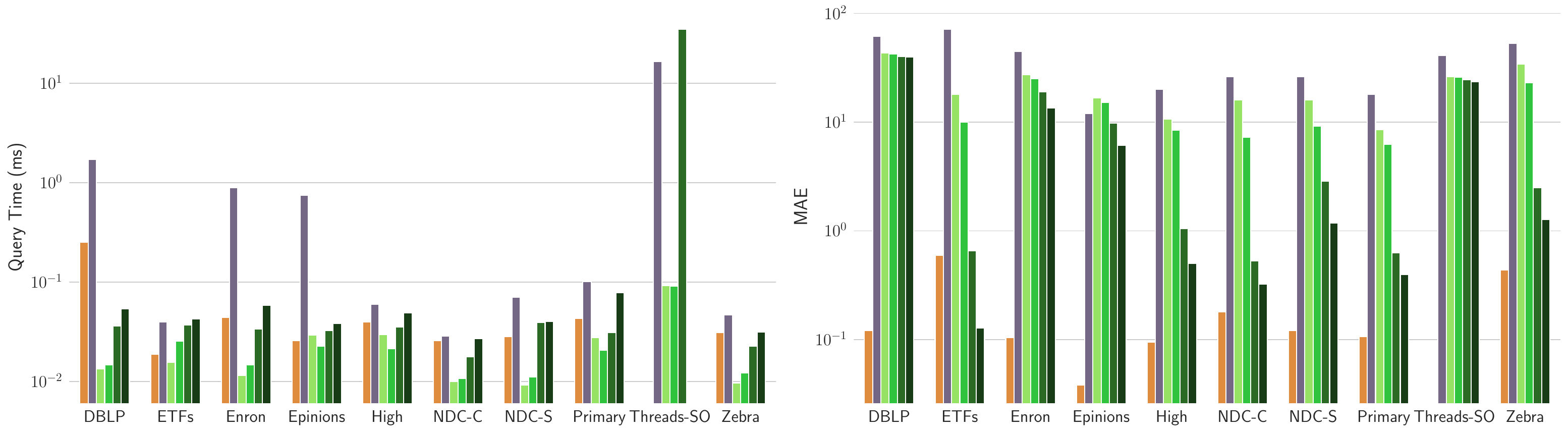}
	\end{tabular}
	\caption{OFF time (upper left), oracle disk size (upper right), TimeXQ (lower left), and MAE (lower right) of \ctl, \hl, and our algorithm varying $\ell$ (\algo-$\ell$) \rev{and fixing \textsf{degree} as landmark selection strategy, \samp as landmark assignment strategy, $\alpha=0.2$, $\beta=0.6$, and $\dmin = 4$}.}
	\label{fig:comparison_ctl_hl}
%\vspace{-3mm}
\end{figure*}

\subsection{Comparison with line-graph-based Oracles}\label{subsec:comparison2}
We build line-graph-based oracles (\Cref{sec:linegraph}) via two state-of-the-art exact oracles for graphs, \ctl~\cite{li2020scaling} and \hl~\cite{farhan2018highly} on top of the line graph.
Note that \hl and \ctl are exact only for connected graphs: \hl exploits this property in the construction of the highway labeling index, while \ctl exploits it in determining core and peripheral nodes.
As a consequence, when the graph is not connected (as often happens in our setting for larger values of $s$), they may introduce errors.

We recall that no existing work can simultaneously answer the three types of $s$-distance queries considered in this work, nor build distance profiles as defined in Problem \ref{prob:dp}.
Thus, we compare our framework against \hl and \ctl, only in the task of hyperedge-to-hyperedge $s$-distance query answering, for $s$ up to $10$.
To this aim, we sampled $100$ pairs of hyperedges $5$ times (for a total of $500$ queries for each $s$), ensuring that \textbf{(i)} most of the pairs are connected for each $s$, \textbf{(ii)} all pairs are $1$-connected, and \textbf{(iii)} most of the pairs belong to (the same) large connected components.
This way, we guarantee that a substantial amount of queries has a finite answer.

\Cref{fig:comparison_ctl_hl} illustrates the metrics of \ctl, \hl, and our algorithm.
The OFF time of \ctl and \hl includes the time required to find the line graph of the hypergraph.
We report the running times of \ctl and \hl for completeness, even though they are not fairly comparable with those of our algorithm because they are written in C++.
We do not show results for \ctl in Threads-SO because the algorithm runs out of memory and is not able to complete.

\spara{OFF.}
For \ctl and \hl we present the creation times with respect to only one parameter combination, because the index creation time is determined principally by the time required to find the line graph.
Hence the total creation time changes only slightly with the parameters of the algorithms, $d$ and $l$.

Instead, we assess the scalability of \algo by varying the parameter $\ell$.
Note that the time required to populate the landmark-based distance oracle depends on several factors: hypergraph size, hypergraph dimension, and hyperedge overlap.
As an example, even though Primary is larger than NDC-S, the running times are comparable because the dimension of NDC-S is larger.
Similarly, IMDB has twice as many hyperedges as Enron and Epinions, but
its dimensionality is less than $\sfrac{1}{6}$ of theirs.
In addition, the average number of hyperedges to which a node belongs is only $2$ in IMDB, but grows to $7.16$ and $18$ in Enron and Epinions.
The larger this number, the more the hyperedges overlap, and thus the more complex the tasks of finding connected components and shortest paths.
The complex behavior of the algorithm is more evident in DBLP. This dataset is one order of magnitude larger than Enron, but the running time is not significantly higher.
This result is again due to the lower dimensionality and average node occurrence of the former dataset.

\spara{Oracle size.}
Especially for larger hypergraphs, the size required to store the \ctl index is much larger than that of storing the landmarks-based oracle.
As an example, the \ctl index for DBLP takes up to 93GB ($d=10$), while \algo takes up to 3.1GB ($\ell=200$).
Note that there is little to no control over the index size of \ctl, while the index size of \algo can be decided by the user according to their needs.
The size of \ctl does not change much with $d$, as the tree-index size is proportional to $d$ and the core-index size is inversely proportional to $d$.
Instead, the index size of \hl increases linearly with $l$, but its total disk size is dominated by the size of the graph, which is required at runtime to compute the distances.

\spara{TimeXQ.}
We evaluate the time required to answer distance queries for 100 pairs of hyperedges, selected with a stratified sampler that ensures a fair amount of finite $s$-distances for each $s \in [1, 10]$.
We repeat the test 3 times and report the average query time.

The TimeXQ of \ctl and \algo is usually lower than that of \hl, with \algo achieving the lowest query times on the largest datasets (DBLP and Enron).
The query time of \algo grows with $\ell$ (more landmarks means more candidate lower and upper bound to consider to get the final estimate), while the query time of \ctl and \hl remains relatively stable when increasing $d$ and $l$.
For \ctl, this result stems from the amount of information stored in the index being basically constant; while for \hl, from the query time being dominated by the time required to evaluate the queries on the sparsified version of the graph.

\spara{MAE.}
\ctl and \hl are exact algorithms designed for connected graphs. As a consequence, in disconnected graphs such as s-line graphs, they are not guaranteed to provide exact answers.
Nevertheless, thanks to a more substantial index, \ctl displays lower MAE values, with higher errors in ETFs and Zebra (despite them being neither the largest nor the most disconnected graphs).
In contrast, \hl, with an index size comparable to that of \algo-200, has the worst performance in all the datasets.

For the largest datasets, there is no significant improvement when increasing $\ell$ for \algo.
Indeed, the number of additional landmarks introduced, and hence additional distance pairs stored, is relatively small compared to the total number of hyperedges.
Thus, the probability of a landmark being on a shortest path between two query hyperedges does not increase much.

\spara{Discussion.}
We reported a comparison on hyperedge-to-hyperedge queries between line-graph-based oracles (\Cref{sec:linegraph}), by using two state-of-the-art methods for connected graphs, and the landmark-based \algo framework (\Cref{sec:solution}).
We cannot compare on \rev{vertex-to-vertex} and \rev{vertex-to-hyperedge} queries, because \ctl and \hl are not designed to work on the augmented line graph.
In particular, since they lack the ability to distinguish nodes representing vertices from nodes representing hyperedges, they cannot ensure to \emph{(i)} perform BFSs traversing only hyperedges, and \emph{(ii)} to select only hyperedges as landmarks.
In the experiments on larger datasets, \algo allows the creation of indices far smaller than those created by \ctl on the line graph, while guaranteeing a small TimeXQ, and MAE comparable with \hl.
\rev{As an example, the \ctl index size for DBLP is $93$GB, whereas our oracle size is $3.1$GB. In addition, we remark that \ctl failed to generate a index for the largest dataset, due to memory issues.}
\rev{We also remark that the query times of \ctl and our oracle may not be directly comparable, primarily due to the fact that \ctl is implemented in C++ while our algorithm is in Java.}
As discussed in \Cref{sec:linegraph}, the limits of line-graph-based oracles emerge with larger hypergraphs, where the line graph size and creation time depend on the characteristics of the hypergraph, and thus can not be controlled.
Instead, the landmark-based \algo framework allows fine-grained control of the trade-off between index size and approximation error at creation time.

\rev{Finally, we highlight that while \hl requires access to the input hypergraph to answer the distance queries, \algo does not.
This characteristic can be advantageous in scenarios where end-users are granted limited access to the original data and are restricted to interacting solely with the oracle.
Nevertheless, in instances where the original hypergraph is accessible, \algo can be adapted to incorporate this information, with the advantage of being able to calculate the exact $s$-distances for pairs of hyperedges within the same small $s$-connected components through bidirectional BFS. This would be, however, at the price of higher query times.}

\subsection{Hypergraph-based Recommendations}\label{sec:casestudy}

\rev{Collaborative filtering recommender systems typically model users as sets of items they liked and each item as the set of users it is liked by.
Hypergraphs, going beyond the binary relations, allows to model relations among sets, 
paving the way for higher-order recommender systems capable of fully leveraging the available information.
The integration of higher-order distance oracles can enhance existing random-walk-based methods~\cite{gori2006research,cooper2014random,gori2007itemrank} designed to assign scores to items that may be of interest to a user.
For instance, if the hypergraph models a user's in-session listened songs, we can recommend a new playlist comprising diverse yet related songs based on the user's previous listening session. In contrast, representing the data as a graph loses the information about which songs were listened to together by the user.}

Following \cite{potamias2009fast}, \rev{in this section, we show how $s$-distance oracles can be embedded into a hypergraph-based recommender system.
One of the advantages of such type of recommender system is that,}
by using higher values of $s$, we can tune the strength of the relations behind the recommendations given.
We consider two labeled hypergraphs: IMDB and DBLP.
In IMDB, each hyperedge represents a movie and each vertex is a member of its crew.
Vertices are associated with a label indicating their role in the movie (e.g., actor, actress, producer, director).
In DBLP, each hyperedge represents a scientific publication and each vertex is an author.
Hyperedges are associated with a label that indicates the venue where the article was published.
The idea is to recommend the $k$-nearest neighbors among the elements which have the same label.
That is, in IMDB we recommend other individuals who have covered the same role of the query vertex and have participated in movies that are close to the ones of the query vertex.
In DBLP, given an article (a hyperedge), we recommend other articles which have been published in the same venue and are close to the query hyperedge in terms of co-authors.
We use \textsc{HypED} to quickly approximate $s$-distances: in the case of IMDB, the oracle answers vertex-to-vertex distance queries, while in the case of DBLP, it answers hyperedge-to-hyperedge distance queries.
As queries, we select 5 labels at random, then we sample vertices or hyperedges with one of those labels.
For comparison, we also compute the exact $k$-nearest neighbors, via BFS.

\begin{table}[t!]
\caption{Avg precision ($\ell = 70$) in finding the top-$k$ closest elements with the same label of the query. }
\vspace{-3mm}
\label{tbl:recommending}
	\centering
	\small
	\begin{tabular}[t]{rcccc}
	\toprule
	&  \multicolumn{2}{c}{AveP@3} &  \multicolumn{2}{c}{AveP@7}\\
	\cmidrule(lr){2-3} \cmidrule(lr){4-5}
	\textbf{s} & IMDB & DBLP & IMDB & DBLP\\
	\midrule
	 2 &   0.699 & 0.587 & 0.760 &   0.619 \\
	 3 &   0.886 & 0.812 & 0.916 &   0.812 \\
	 4 &   0.967 & 0.927 & 0.980 &   0.927 \\
	 5 &   0.988 & 1.000 & 0.995 &   1.000 \\
	 6 &   0.993 & 1.000 & 0.996 &   1.000 \\
	 7 &   0.997 & 1.000 & 1.000 &   1.000 \\
	 8 &   1.000 & 1.000 & 1.000 &   1.000 \\
	 9 &   1.000 & 1.000 & 1.000 &   1.000 \\
	10 &   1.000 & 1.000 & 1.000 &   1.000 \\
	\bottomrule
	\end{tabular}
	\vspace{-5mm}
\end{table}

\spara{Precision@k.} \Cref{tbl:recommending} reports the average precision in the computation of the top-$3$ and top-$7$ closest elements to the query, varying $s$.
With such a small oracle size, the MAE of \textsc{HypED} is $0.90$ in DBLP, and $1.44$ in IMDB.
Nevertheless, we can see that for $s \geq 3$ the average precision is close to optimal, i.e., the rankings provided by \textsc{HypED} are almost identical to the exact ones.while being produced in a fraction of the time required by an exact $k$-nearest neighbors computation.

\begin{table*}
\caption{Top-5 suggestions provided by \textsc{HypED} for a query article, for two values of the parameter $s$.}
\label{tbl:recommending2}
	\centering
	\footnotesize
\resizebox{\textwidth}{!}{
	\begin{tabular}[t]{c}
		\toprule
		\multicolumn{1}{c}{\textbf{Query:} \emph{Incorporating Pre-Training in Long Short-Term Memory Networks for Tweets Classification.}}\\
		\midrule
	 \textbf{Top-5 Suggestions $(s = 1)$}\\
		\midrule
		Personalized Grade Prediction: A Data Mining Approach\\
		Predicting Sports Scoring Dynamics with Restoration and Anti-Persistence\\
		R2FP: Rich and Robust Feature Pooling for Mining Visual Data\\
		Mining Indecisiveness in Customer Behaviors\\
		New Probabilistic Multi-graph Decomposition Model to Identify Consistent Human Brain Network Modules\\
			\midrule
	 \textbf{Top-5 Suggestions $(s = 2)$}\\
	\midrule
		R2FP: Rich and Robust Feature Pooling for Mining Visual Data\\
		New Probabilistic Multi-graph Decomposition Model to Identify Consistent Human Brain Network Modules\\
		KnowSim: A Document Similarity Measure on Structured Heterogeneous Information Networks\\
		Leveraging Implicit Relative Labeling-Importance Information for Effective Multi-label Learning\\
		Generative Models for Mining Latent Aspects and Their Ratings from Short Reviews\\
		\bottomrule
	\end{tabular}
}
\vspace{-2mm}
\end{table*}

\spara{Anecdotal Evidence.}
\Cref{tbl:recommending2} shows an example of the top-5 suggestions for a query (an article) in DBLP, for two values of $s$.
The query considered is the article \emph{`Incorporating Pre-Training in Long Short-Term Memory Networks for Tweets Classification'}.
The fields of research of the authors are data mining, machine learning, and deep learning.
Different values of $s$ bring different sets of recommendations and different rankings.
In particular, the article \emph{`Predicting Sports Scoring Dynamics with Restoration and Anti-Persistence'} takes the second position in the rank by $1$-distance, but its $2$-distance from the query is so large that it is not included in the rank by $2$-distance.
This result indicates that the two articles are weakly related, as it can be gathered by examining the topics they address.
The two articles \emph{`R2FP: Rich and Robust Feature Pooling for Mining Visual Data'} and \emph{`New Probabilistic Multi-graph Decomposition Model to Identify Consistent Human Brain Network Modules'} are included in both rankings, but the second article is in a higher position in the rank by $2$-distance.
Clearly, the article is farther from the query, but its connection is stronger.
Finally, by looking at the topics studied in the query and in the suggestions per ranking, we can see that the article \emph{`KnowSim: A Document Similarity Measure on Structured Heterogeneous Information Networks'} overlaps more with the query, and hence it may be more interesting for the user.
In fact, the query article proposes a deep learning algorithm that learns dependencies between words in tweets to solve a tweet classification task, while KnowSim builds heterogeneous information networks to identify similar documents (e.g., tweets), and hence can effectively classify documents into any desired number of classes.

As another example, in IMDB, given as query the producer Barry Mendel, our method suggests the producer Yuet-Jan Hui (based vertex-to-vertex distance), together with
the movie \emph{``The Eye''} (vertex-to-hyperedge distance), that is the closest movie to Barry Mendel produced by Yuet-Jan Hui.
These suggestions seem legitimate, as Mendel is known for \emph{``The Sixth Sense''}, a mystery drama featuring a boy able to communicate with spirits, while \emph{``The Eye''} is a mystery horror featuring a violinist who sees dead people.

\subsection{Approximate s-closeness Centrality}\label{sec:centrality}
It has been shown that well-connected proteins in PPI networks are more likely to play crucial roles in cellular functions~\cite{jeong2001lethality}.
Thus, several works have proposed to use centrality measures to detect relevant genes and drug targets~\cite{soofi2020centrality,kotlyar2012network,joy2005high,viacava2021centrality}.

We evaluate the ability of \textsc{HypED} to approximate the $s$-closeness centrality of vertices and hyperedges in the four PPI hypergraphs in \Cref{tbl:datasets}.
For a hyperedge $e$ in a $s$-connected component $c_s$, the $s$-closeness centrality of $e$ is $cl(e, s, c_s) = \sum_{f \in c_s}\Delta_{s}(e, f) / (|c_s| - 1)$.
For a vertex $v$ belonging to a set of hyperedges $e_1, \dots, e_p$ in the $s$-connected components $c_s^1, \dots, c_s^p$, the $s$-closeness centrality of $v$ is $\max_j{cl\left(e_j, s, c_s^j\right)}$.\footnote{In contrast to hyperedges, for $s > 1$, a vertex $v$ may belong to different $s$-connected components, as it may be in hyperedges that overlap only in $v$.}
We use \emph{max} to give more importance to vertices appearing only in large $s$-connected components than to those belonging to many trivial $s$-connected components, although other aggregation functions can be used.

\Cref{fig:centrality} reports the MAPE and LAR for \textsc{HypED}-$100$ in computing the $s$-closeness centrality of $100$ random vertices (v) and hyperedges (h) in each PPI dataset.
Both are measures of relative accuracy: MAPE measures the mean absolute percentage error, while LAR~\cite{tofallis2015better} is the sum of squares of the log of the accuracy ratio.
Due to the hypergraph fragmentation at larger $s$, vertices likely belong to small $s$-connected components.

Thanks to the procedure used by \textsc{HypED} to approximate the $s$-distances in small components, the centrality of such vertices is well approximated (vertex centrality is defined as a \emph{max}).
Thus, we observe smaller errors for vertices than hyperedges.
When increasing $s$, the error usually increases, because the hypergraph becomes more fragmented.
The error is low in all but the chem-gene dataset, hence proving the ability of \textsc{HypED} in accurately approximating the $s$-closeness centrality of both vertices and hyperedges, in the case of higher-order PPIs.
The higher error in chem-gene is due to its higher-order organization, and in particular, to the presence of a big $s$-connected component with few hubs, and many vertices with low centrality.
In such cases, more landmarks are needed to better approximate the centrality, as each landmark is likely to be on only a few shortest $s$-paths.

\begin{table}[t!]
	\centering
	\small
	\caption{MAPE and LAR for \textsc{HypED}, in computing the centrality of vertices (v) and hyperedges (h) in the PPI datasets.}
	\label{fig:centrality}
%\vspace{-3mm}
	\sisetup{table-format=2.4, table-alignment-mode=format}
	\begin{tabular}{l ccccc}
		\toprule
		& & \multicolumn{2}{c}{v} & \multicolumn{2}{c}{h}\\
		\cmidrule(lr){3-4} \cmidrule(lr){5-6}
		\textbf{Dataset} & \textbf{s} & MAPE & LAR & MAPE & LAR\\
		\midrule
		\multirow{2}{*}{chem-gene}
			& 2 & 0.6077 & 0.2540 & 0.5645 & 0.2218\\
		 	& 3 & 0.5929 & 0.2557 & 0.3866 & 0.1230\\
		\hline
		\multirow{2}{*}{dis-function}
			& 2 & 0.0098 & 0.0027 & 0.1376 & 0.0292\\
			& 3 & 0.0107 & 0.0020 & 0.2181 & 0.0567\\
		\hline
		\multirow{2}{*}{dis-gene}
			& 2 & 0.0840 & 0.0065 & 0.1795 & 0.0347\\
			& 3 & 0.0474 & 0.0021 & 0.0986 & 0.0130\\
		\hline	
		\multirow{2}{*}{dis-chemical}
		 	& 2 & 0.0096 & 0.0002 & 0.2610 & 0.0753\\
			& 3 & 0.0107 & 0.0002 & 0.2857 & 0.0868\\
		\bottomrule
	\end{tabular}
\end{table}

% !TEX root = ../main.tex
\section{Conclusions and future work} This paper introduces \algo, an oracle for point-to-point distance queries in hypergraphs.
Our definition of distance includes a parameter $s$ to tune the required level of overlap for two hyperedges to be considered adjacent.
Higher values of $s$ emphasize stronger relations between the entities.
This strength comes with additional challenges, which require targeted solutions beyond the application of general-purpose graph-based algorithms.

We first presented an oracle based on $(i)$ creating an augmented line graph of the hypergraph, and $(ii)$ building on top of it any state-of-the-art distance oracle for standard graphs. We discussed several important limitations of this approach, and thus proposed a method based on landmarks to produce an accurate oracle with a predefined size directly on the hypergraph.
As a side---yet important---contribution, we devised an efficient algorithm for computing the $s$-connected components of a hypergraph.
The experimental comparison with state-of-the-art exact solutions showed that our framework \textbf{(i)} can answer queries as efficiently, even if they trade space for time; and \textbf{(ii)} can be constructed as efficiently, even though it requires finding all the $s$-connected components of the hypergraph.

\emph{To the best of our knowledge, this is the first work to propose point-to-point distance query oracles in hypergraphs, as well as to propose an algorithm for computing $s$-connected components in hypergraphs.}

\rev{Our framework distinguishes itself by handling vertex-to-vertex $s$-distance queries, a key aspect in measuring proximity within a hypergraph while considering the strength of relations between hyperedges.
This is exemplified in scenarios such as song recommendation, where higher values of $s$ lead to more targeted suggestions based on direct connections, while lower values of $s$ promote serendipity by exploring more indirect links between songs.
By looking at the top-$k$ $s$-reachable vertices from a given song, we can offer diverse yet consistent song recommendations.}

\rev{We emphasize the distinction between vertex-to-vertex $s$-distance in the hypergraph and hyperedge-to-hyperedge $s$-distance in its dual hypergraph.
An $s$-path in the hypergraph is a sequence of hyperedges with consecutive hyperedges sharing at least $s$ common vertices, whereas an $s$-path in the dual hypergraph is a sequence of vertices with consecutive vertices belonging to at least $s$ common hyperedges.
As a result, vertices that are $s$-connected in the hypergraph may not be $s$-connected in the dual hypergraph.
Both definitions hold relevance and may cater to different requirements of the specific application scenario.
The ability of our framework to compute both types of distances provides users with a comprehensive toolset for making informed decisions based on their specific objectives.}

The current work is just a first step in a new area and opens several avenues for further investigation.
Similar to the wide literature on distance oracles in graphs, new different methods can be tried for the same problem we introduced, and specialized approaches can be devised for hypergraphs having specific characteristics, or emerging in specific application domains.

For instance, one could consider hypergraphs where each hyperedge has the same cardinality, or, in dynamic settings, the problem of maintaining the oracle incrementally as the underlying hypergraph evolves.

\bibliographystyle{plain}
\bibliography{ref}

\end{document}